\documentclass[lettersize,journal]{IEEEtran}
\usepackage{amsmath,amsfonts}
\usepackage{algorithmic}
\usepackage{algorithm}
\usepackage{array}
\usepackage[caption=false,font=normalsize,labelfont=sf,textfont=sf]{subfig}
\usepackage{textcomp}
\usepackage{stfloats}
\usepackage{url}
\usepackage{verbatim}
\usepackage{graphicx}
\usepackage{cite}
\usepackage{array}
\usepackage{multirow}
\usepackage{tabularx,colortbl}
\usepackage[dvipsnames]{xcolor}
\usepackage{lipsum}
\definecolor{lavender}{HTML}{E6E6FA}
\newcolumntype{C}[1]{>{\centering\let\newline\\\arraybackslash\hspace{0pt}}m{#1}}
\usepackage{xcolor}
\usepackage{pifont}
\let\oldding\ding
\renewcommand{\ding}[2][1]{\scalebox{#1}{\oldding{#2}}}
\usepackage[caption=false]{subfig}
\newcommand{\ineq}[1]{\footnotesize$#1$\normalsize}{}

\newcommand{\tech}{QUANTISENC}

\begin{document}

\title{\huge \tech{}: A Fully-Configurable Software-Defined Digital Quantized Spiking Neural Core Architecture}
\title{\huge An Open-Source Fully-Configurable Software-Defined \\Digital Quantized Spiking Neural Core Architecture}
\title{\huge A Fully-Configurable Open-Source 
 Software-Defined Digital Quantized Spiking Neural Core Architecture}

\author{Shadi Matinizadeh, Noah Pacik-Nelson, Ioannis Polykretis, Krupa Tishbi, Suman Kumar, M. L. Varshika, Arghavan Mohammadhassani, Abhishek Mishra, Nagarajan Kandasamy, James~Shackleford, Eric Gallo, Anup~Das
\thanks{S. Matinizadeh, K. Tishbi, S. Kumar, M. L. Varshika, A. Mohammadhassani, A. Mishra, N. Kandasamy, J. Shackleford, and A. Das are with the Department
of Electrical and Computer Engineering, Drexel University, Philadelphia,
PA, 19104 USA. Correspondance e-mail: anup.das@drexel.edu.}
\thanks{N. Pacik-Nelson, I. Polykretis, and E. Gallo are with Accenture Labs.}
}

\markboth{arXiv}%
{Shadi and Das \MakeLowercase{\textit{et al.}}: A Fully-Programmable Digital Quantized Spiking Neural Core Architecture}


\maketitle

\begin{abstract}
We introduce \tech{}, a fully configurable open-source software-defined digital quantized spiking neural core architecture to advance research in neuromorphic computing.
\tech{} is designed hierarchically using a bottom-up methodology with multiple neurons in each layer and multiple layers in each core.
The number of layers and neurons per layer can be configured via software in a top-down methodology to generate the hardware for a target spiking neural network (SNN) model.
\tech{} uses leaky integrate and fire neurons (LIF) and current-based excitatory and inhibitory synapses (CUBA).
The nonlinear dynamics of a neuron can be configured at run-time via programming its internal control registers.
Each neuron performs signed fixed-point arithmetic with user-defined quantization and decimal precision.
\tech{} supports all-to-all, one-to-one, and Gaussian connections between layers.
Its hardware-software interface
is integrated with a PyTorch-based SNN simulator.
This integration allows to define and train an SNN model in PyTorch and evaluate the hardware performance (e.g., area, power, latency, and throughput) through FPGA prototyping and ASIC design.
The hardware-software interface also takes advantage of the layer-based architecture and distributed memory organization of \tech{} to enable pipelining by overlapping computations on streaming data.
Overall, the proposed software-defined hardware design methodology
offers flexibility similar to that of high-level synthesis (HLS), but provides better hardware performance with zero hardware development effort. 
We evaluate \tech{} using three spiking datasets and show its superior performance against state-of-the-art designs. 
\end{abstract}


\section{Introduction}\label{sec:intro}
\IEEEPARstart{N}{euromorphic} computing describes the hardware implementation of the biological nervous system~\cite{mead1990neuromorphic}.
Over the years, architects and designers have created such systems using analog and digital components, 
offering lower power consumption, reduced latencies, and many other benefits seen in biological systems.
Although analog designs improve energy efficiency by taking advantage of electronic and physical laws in implementing neurons and synapses, digital designs are faster to implement on silicon due to the maturity of their design flow, while
benefiting from technology scaling~\cite{joubert2012hardware}. 

Existing digital neuromorphic designs suffer from the following three key limitations, which we address in this work.
First, once implemented on silicon, existing designs allow only programming of its synaptic weights. 
They do not allow configuring the neuron dynamics at run-time.
We show that run-time reconfiguration can be used to explore the trade-off between performance and power (Section~\ref{sec:dynamic}). 
Second, existing designs lack a well-defined hardware-software interface to stream input, visualize output, and load design configurations and synaptic weights.
We show that a hardware-software interface can also be used to allow pipelining to improve performance (Section~\ref{sec:throughput}).
Finally, existing neuromorphic designs are not released to the community.

\begin{figure*}[b!]
    \centering
    \centerline{\includegraphics[width=1.5\columnwidth]{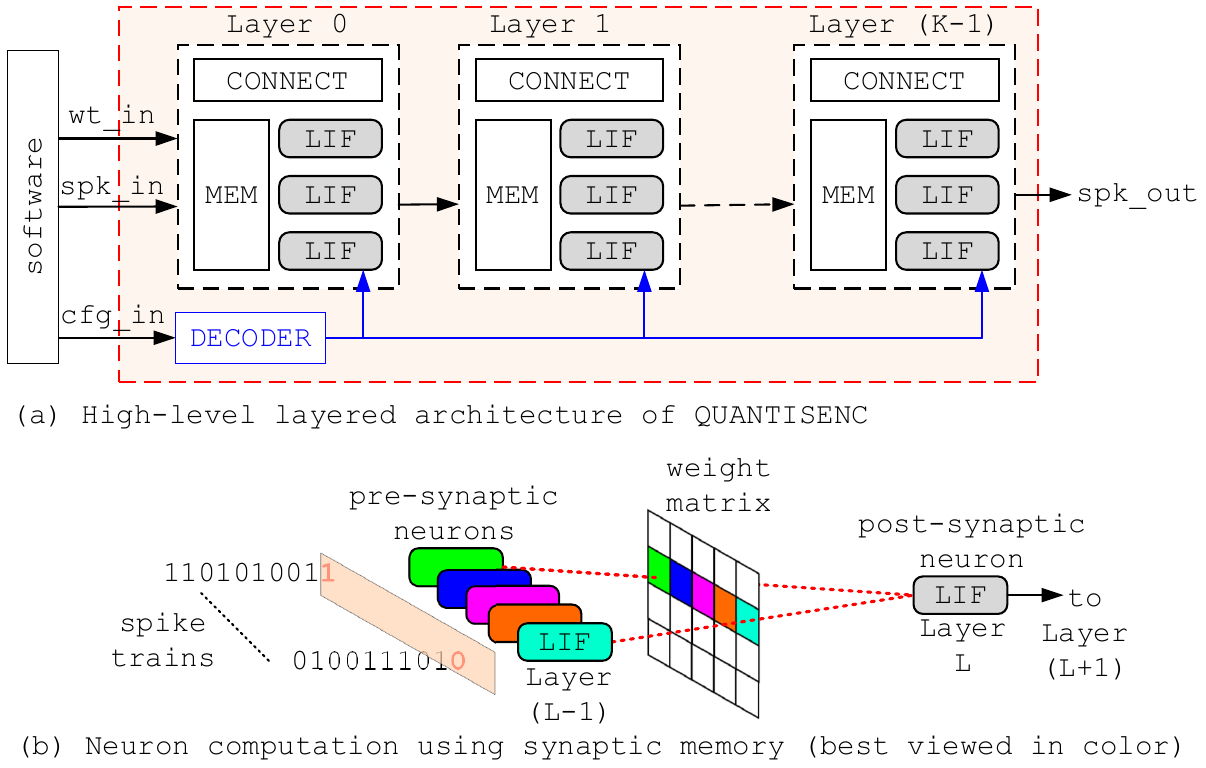}}
    \caption{(a) High-level overview of \tech{'s} design blocks. (b) Computation of spike trains using synaptic weights.}
    \label{fig:snncore}
\end{figure*}

We introduce \tech{} (QUANTIzed Spiking-Enabled Neural Core), a fully configurable layer-based neuromorphic hardware design and its software-defined hardware design methodology 
for spiking neural networks (SNNs)~\cite{maass1997networks}. 
This design methodology allows researchers and system designers to train an SNN model in software (e.g., Python) and evaluate its hardware performance, such as area, power, latency, and throughput through FPGA prototyping and ASIC design.
The following are our key \textbf{contributions}.

\IEEEpubidadjcol

\begin{enumerate}
    \item \textbf{Configurability:} 
    \tech{} is designed hierarchically using a bottom-up methodology with neurons, layers, and cores. The number of layers and neurons per layer can be configured by software in a top-down methodology to generate hardware for a target SNN model.
    \tech{} supports signed computations of fixed-point arithmetic with user-defined quantization and decimal precision.
    \tech{} implements leaky integrate-and-fire neurons (LIF)~\cite{indiveri2003low}, and current-based excitatory and inhibitory synapses (CUBA), with all-to-all, one-to-one, and Gaussian connections between layers. 
    Due to our modular design methodology, \tech{} can be easily extended to support other types of neurons, e.g., Izhikevich and compartmental~\cite{ward2022beyond}, and synapse, e.g., conductance-based synapse (COBA)~\cite{cessac2008dynamics}.
    
    \item \textbf{Hardware-Software Interface:}
    A key novelty of our design methodology is that it allows the LIF dynamics to be configured from an application software (e.g., PyTorch code) at runtime by programming the internal control registers on the hardware through its interface with the system software (e.g., OS).
    This provides flexibility to fine-tune hardware performance after the design has been fabricated on silicon.
    Our hardware-software interface is also used to program synaptic weights, drive input, and visualize hardware output.
    \item \textbf{Pipelining:} \tech{} implements a distributed memory organization, where the synaptic memory of neurons in a layer is instantiated within each layer, allowing the layers of \tech{} to operate independently and parallel to each other.
    Although \tech{} is a dataflow architecture with layer-by-layer processing of input data, we leverage the parallelism in its internal layer design by overlapping the processing of streaming data in a pipelined fashion using the proposed hardware-software interface.
    This refinement of dataflow computations significantly improves throughput.

    \item \textbf{Open-Source Release:} \tech{} is developed over a period of eight years and is supported by Accenture LLP, three US National Science Foundation research grants, one US Department of Energy grant, and one European Horizon 2020 research grant.
    \tech{} is developed to foster future research in neuromorphic computing.
    To obtain a copy of the design, please email Shadi Matinizadeh at sm4884@drexel.edu.
\end{enumerate}

We evaluate \tech{} and our system design methodology using Spiking MNIST~\cite{fatahi2016evt_mnist}, DVS Gesture~\cite{amir2017low}, and Spiking Heidelberg Digit (SHD)~\cite{cramer2020heidelberg} datasets.
Our results show that compared to the software implementation, the hardware implementation of LIF dynamics has a root mean square error (RMSE) of only 0.25 mV using a quantization of 16 bits with a decimal precision of 7 bits. 
The RMSE increases to 0.43 mV, when using a quantization of 8 bits with a decimal precision of 3 bits. 
For datasets, \tech{} significantly outperforms state-of-the-art designs in terms of resource utilization, power, latency, and throughput. 
We show the scalability of \tech{} for different datasets and for the settings of its design parameters.

\section{High-Level Overview of \tech{}}\label{sec:overview}
Figure~\ref{fig:snncore}a shows a high-level overview of the proposed single-core design \tech{}.
There are \ineq{K} layers, each instantiating \texttt{LIF} neuron modules.
The number of layers and neurons per layer are design parameters that can be configured to implement a target SNN model in hardware.
LIF parameters are stored in the control registers of the \texttt{decoder} module.
Within each layer, the \texttt{connect} module consists of connection parameters
to specify the organization of its synaptic weight memory (\texttt{MEM}). 

\begin{figure*}[b!]
    \centering
    \centerline{\includegraphics[width=1.99\columnwidth]{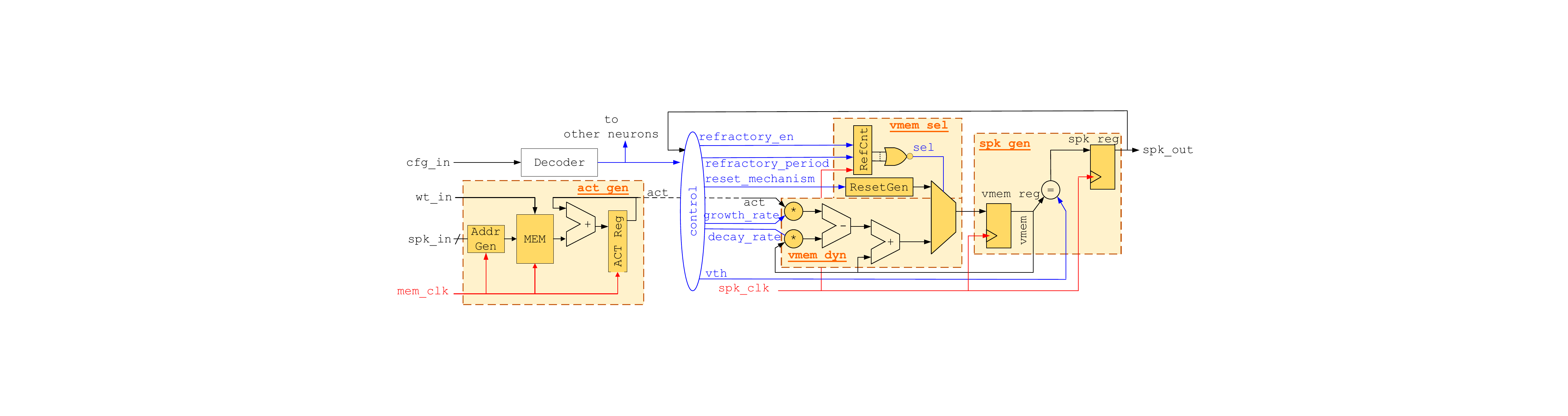}}
    \caption{Design of a leaky-integrate-and-fire (LIF) neuron. There are four main design components -- \texttt{ActGen}, \texttt{VmemDyn}, \texttt{VmemSel}, and \texttt{SpkGen}.}
    \label{fig:lif}
\end{figure*}

At a high level, the synaptic memory of \tech{} is distributed among its layers such that all pre-synaptic weights
are stored in their respective layer.
This distributed memory organization allows us to pipeline the design to improve throughput.
Without loss of generality, we illustrate a fully connected layer in Figure~\ref{fig:snncore}b, where a neuron in layer \ineq{L} (post-synaptic neuron) receives input from all neurons in layer \ineq{(L-1)} (pre-synaptic neurons).
The output spikes of this neuron are routed to all neurons in the layer \ineq{(L+1)}.

At a finer granularity, synaptic memory (\texttt{MEM}) 
is a \ineq{M\times N} weight matrix. This weight matrix corresponds to a design configuration of \ineq{N} neurons in the layer with \ineq{M} pre-synaptic connections per neuron.
Spike trains from pre-synaptic connections are weighed using synaptic weights to generate output spike trains as illustrated in the figure.

The access granularity of a layer's synaptic memory is that of a synaptic weight between a pair of pre-synaptic and post-synaptic neurons.
Therefore, each weight in \tech{} can be addressed and programmed individually.

The I/O interface to \tech{} is as follows.

\begin{itemize}
    \item \texttt{wt\_in:} This interface is used to program synaptic memory by specifying weight addresses and data. 
    This interface is clocked using \texttt{mem\_clk}.
    \item \texttt{cfg\_in:} This interface is used to configure the neuron parameters and the neuron dynamics. 
    We implement control registers inside the \texttt{decoder} module to store these parameters.
    These registers are clocked using \texttt{mem\_clk}.
    \item \texttt{spk\_in/out:} This interface is used for data input and output of \tech{}.  
    This interface is clocked using \texttt{spk\_clk}, which is the main design clock. Each spike is encoded using the address event representation (AER) format, which is a standard representation technique for spikes in the digital domain. 
\end{itemize}

The I/O interface of \tech{} is through the system software, which we elaborate in Section~\ref{sec:software}.

\section{Detailed Implementation of 
 \tech{}}\label{sec:hardware}

\subsection{Neuron Implementation}\label{sec:lif}
At the core of \tech{} are leaky integrate and fire neurons (LIF).
Figure~\ref{fig:lif} shows the design of a digital LIF neuron based on the analog implementation proposed in~\cite{indiveri2003low}.
A neuron dynamic is defined by the first-order ordinary differential equation (ODE) as
\begin{equation}
    \label{eq:lif_ode}
    \footnotesize \tau\frac{dU(t)}{dt} = -U(t) + R\cdot I_{in}(t)
\end{equation}
where \ineq{U(t)} is the membrane potential shown as \texttt{vmem}, \ineq{\tau = R\cdot C} is the time constant of the neuron defined using membrane resistance \ineq{R} and membrane capacitance \ineq{C}, and \ineq{I_{in}(t)} is the input current to the neuron. 
Equation~\ref{eq:lif_ode} can be solved using the forward Euler method as
\begin{equation}
    \label{eq:lif_pre}
    \footnotesize U(t+\Delta t) = U(t) + \frac{\Delta t}{\tau} \Big(-U(t) + R\cdot I_{in}(t)\Big)
\end{equation}
Equation~\ref{eq:lif_pre} can be rewritten using neuron parameters as
\begin{equation}
    \label{eq:lif_post}
    \footnotesize U(t+\Delta t) = U(t) - \texttt{decay\_rate}\cdot U(t) + \texttt{growth\_rate}\cdot I_{in}(t)
\end{equation}
where 
\begin{equation}
    \label{eq:decay}
    \footnotesize \texttt{decay\_rate} = \frac{\Delta t}{\tau} = \frac{\Delta t}{R\cdot C}, \text{ and }
\end{equation}

\begin{equation}
    \label{eq:grow}
    \footnotesize \texttt{growth\_rate} = R\cdot \frac{\Delta t}{\tau} = \frac{\Delta t}{C}
\end{equation}

Decay and growth rates are design parameters that are programmed in the control registers within the decoder module.
The digital implementation of Equation~\ref{eq:lif_post} is shown in Figure~\ref{fig:lif} inside the bounding box marked as \texttt{VmemDyn}.

The input current \ineq{I_{j}(t)} to the neuron \ineq{j} is the neuron activation (\texttt{act}), which is calculated as the weighted sum of input spikes from all its pre-synaptic connections as shown in Equation~\ref{eq:act} for
current-based synapse (CUBA).
\begin{equation}
    \label{eq:act}
    \footnotesize I_{j}^j(t) = \sum_i x_{ij} \cdot w_{ij},
\end{equation}
where \ineq{x_{ij}} is the spike from neuron \ineq{i} to neuron \ineq{j} and \ineq{w_{ij}} is the synaptic weight between these neurons. 
The bounding box \texttt{ActGen} in Figure~\ref{fig:lif} shows the implementation of a neuron activation. 
The address generator operates on \texttt{mem\_clk} to generate memory addresses for all synaptic weights \ineq{w_{ij}}'s. 
The \texttt{act\_reg} is used to store the weighted sum, which is calculated as follows.
For each pre-synaptic connection, the synaptic weight is added to the weighted sum if there is an input spike on the connection.
In general, it takes \ineq{M} \texttt{mem\_clk} cycles to compute the weighted sum for a neuron with \ineq{M} pre-synaptic connections.
This implementation allows synaptic memory to be implemented using dedicated RAM, such as Block RAM (BRAM) or Ultra RAM (URAM) in FPGA~\cite{gungor2022optimizing}.
\tech{} can also be configured to store synaptic weights in Lookup Tables and Flip-flops.
We show the performance trade-off for these design choices in Sec.~\ref{sec:throughput}.

\begin{figure}[h!]
    \centering
    \centerline{\includegraphics[width=0.99\columnwidth]{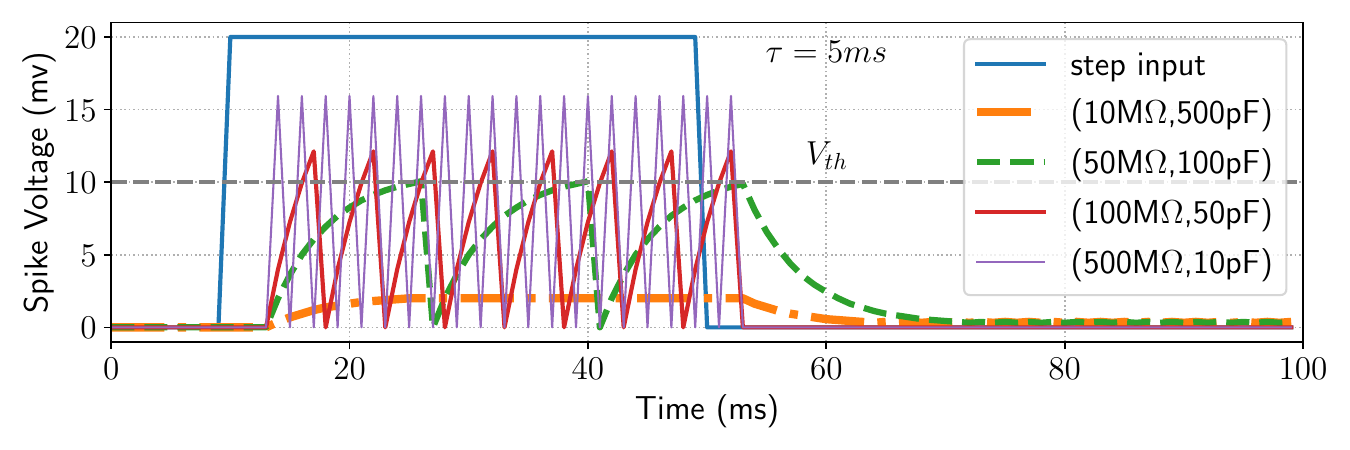}}
    \caption{Impact of $R$ and $C$ settings on a neuron dynamics.}
    \label{fig:rc_50}
\end{figure}

Figure~\ref{fig:rc_50} shows the impact of the \ineq{R} and \ineq{C} settings on membrane dynamics, when a neuron is excited with a step input of 40 ms duration.
We set a membrane time constant \ineq{\tau = 5} ms and a threshold voltage \ineq{V_{th} = 10} mv.
We make the following three key observations.

First, with large \ineq{R} and small \ineq{C} (\ineq{R = 500M\Omega} and \ineq{C = 10pF}), the membrane potential increases significantly due to a large \texttt{growth\_rate} (Equation~\ref{eq:grow}), frequently exceeding the threshold voltage and generating a large number of spikes.
Second, reducing \ineq{R} while increasing \ineq{C} (keeping \ineq{\tau = RC} constant) reduces the \texttt{growth\_rate}, which reduces the number of spikes.
Finally, with a small \ineq{R = 10M\Omega} and a large \ineq{C = 50 0pF}, the membrane potential does not cross the threshold voltage. 
Therefore, no spike is generated from the neuron.
We note that both hardware performance (e.g., power consumption) and application performance (e.g., accuracy) are function of the number of spikes generated from a neuron, which depends on the \ineq{R} and \ineq{C} settings.
\tech{} allows exploring these parameters to achieve the desired performance.

The  \texttt{SpkGen} bounding box in Figure~\ref{fig:lif} generates a spike when the membrane potential \ineq{U(t)} crosses the threshold voltage \ineq{V_{th}}.
After firing a spike, the following two independent processes are triggered in a neuron.
\subsubsection{Reset Mechanism} A membrane potential is reset using one of the following four user-defined reset mechanisms, i.e.,
\begin{equation}
    \label{eq:reset}
    \footnotesize U(t) = \begin{cases}
        V_{reset} & \textbf{Reset-to-Constant}\\
        0 & \textbf{Reset-to-Zero}\\
        U(t) - V_{th} & \textbf{Reset-by-Subtraction}\\
        U(t) - \texttt{decay\_rate}*U(t) & \textbf{Default}
    \end{cases}
\end{equation}

\begin{figure}[h!]
    \centering
    \centerline{\includegraphics[width=0.99\columnwidth]{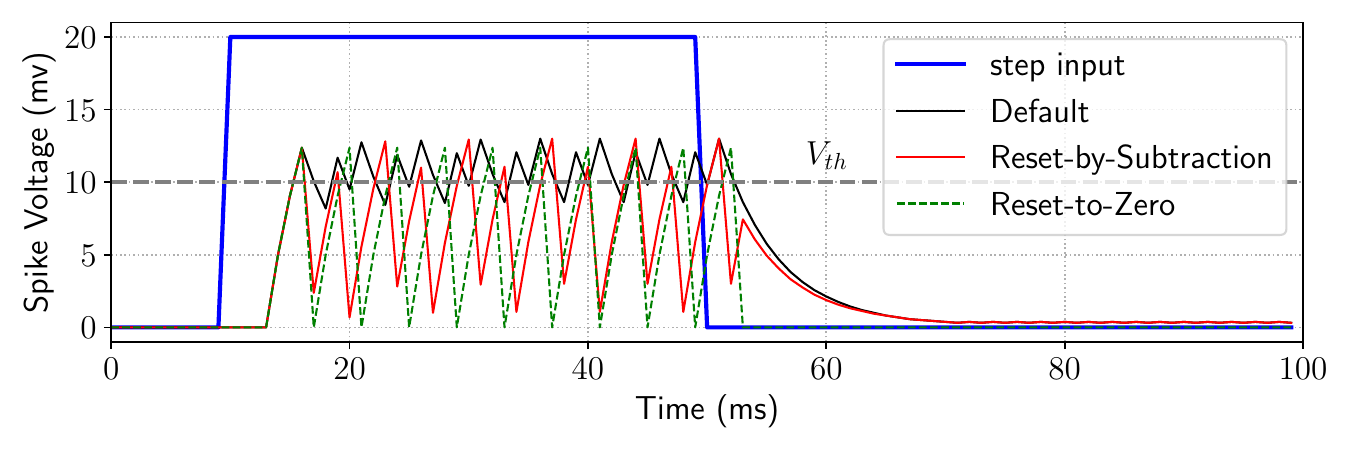}}
    \vspace{-10pt}
    \caption{Neuron dynamics for different reset mechanisms.}
    \label{fig:reset}
\end{figure}

Figure~\ref{fig:reset} shows the reset dynamics of a neuron for a step input of 40 ms. 
The default reset behavior is an exponential decay of the membrane potential.
Without a refractory mechanism (which we describe next), the membrane potential decays only marginally due to continuous excitation, generating a total of 37 spikes within the 40 ms time window.

In the \textit{Reset-by-Subtraction} mechanism, the threshold voltage is subtracted from the membrane potential of a neuron after generating a spike.
This allows the membrane potential to decay considerably after firing a spike, resulting in fewer spikes (= 14) than the default reset mechanism.
Finally, in the \texttt{Reset-to-Zero} mechanism, a neuron membrane potential is reset to zero after firing a spike, requiring additional time for the membrane potential to cross the threshold voltage and generate a spike.
The total number of spikes generated using this mechanism is the least among the three reset mechanisms.

\subsubsection{Refractory Mechanism} A neuron membrane potential is held constant for a user-defined time duration called the \texttt{refractory period} after a neuron fires a spike. 
The refractory mechanism prevents the neuron from firing another spike in the refractory window.
The design of the refractory mechanism consists of a counter (\texttt{RefCnt}), which is activated every time a spike is generated.
Once activated, the counter counts down from the user-defined \texttt{refractory\_period} to zero at every edge of \texttt{spk\_clk}.
During this period of time, the neuron does not generate spikes.
The refractory mechanism regularizes spike trains generated from a neuron by setting an upper limit on the spike firing frequency (\ineq{f_\text{Max}}) as
\begin{equation}
    \label{eq:spike_frequency}
    \footnotesize f_\text{Max} \le \frac{1}{\texttt{refractory\_period}}
\end{equation}

The reset techniques and the refractory mechanism are implemented inside the \texttt{VmemSel} block in Figure~\ref{fig:lif}.

\subsection{Synapse Implementation}\label{sec:connection}
Unlike many previous works in which layers are fully connected to each other, \tech{} supports additional connection modalities, allowing it to implement many modern SNN models. 
To formally introduce these connections, we represent the synaptic weight \ineq{w_{ij}} between neurons \ineq{i} and \ineq{j} as \ineq{w_{ij} = \alpha_{ij}\cdot\beta_{ij}\cdot\omega_{i,j}}, where \ineq{\omega_{ij}} is the absolute synaptic weight. and \ineq{\alpha_{ij}} and \ineq{\beta_{ij}} are weight parameters defined as follows.

\subsubsection{Synaptic Connections (Network Topology)}
Consider a layer \ineq{L} in the design. 
The \textbf{connection parameter} \ineq{\alpha_{ij} \in \{0,1\}} for layer \ineq{L} is defined as
\begin{equation}
    \label{eq:connections}
    \footnotesize \begin{cases}
        a)~~\alpha_{ij} = 1~\forall i\in (L-1)~\&~\forall j \in L & \textbf{: all-to-all} \\
        b)~~\alpha_{ij} = 1~\textbf{ if } i = j, \text{ and } 0 \text{ otherwise} & \textbf{: one-to-one}\\
        c)~~\alpha_{ij} = 1~~\textbf{ if } |i - j| \leq 1, \text{ and } 0 \text{ otherwise} & \textbf{: Gaussian}\\
    \end{cases}
\end{equation}

\begin{figure}[h!]
    \centering
    \centerline{\includegraphics[width=0.99\columnwidth]{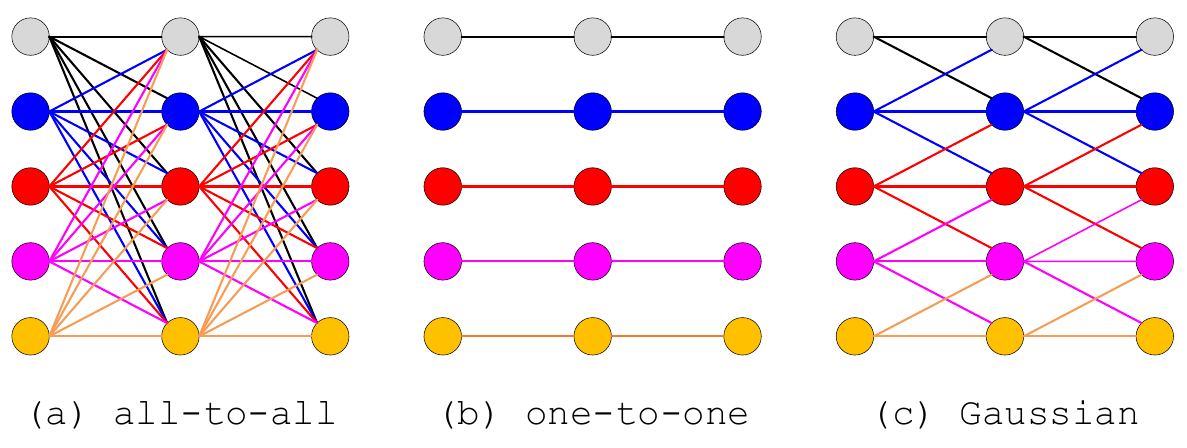}}
    \caption{Connection modalities supported in \tech{}.}
    \label{fig:connections}
\end{figure}

Figure~\ref{fig:connections} shows the three different connection modalities supported in \tech{}. 
For \textbf{all-to-all} connections (also called \textbf{full} connections), each neuron in layer \ineq{L} receives input from every neuron in layer \ineq{(L-1)}. This is defined in Equation~\ref{eq:connections}a and is shown in Figure~\ref{fig:connections}a.
For \textbf{one-to-one} connections, a neuron at a specific index in a layer is only connected to the neuron at the same index in the next layer. 
This is defined in Equation~\ref{eq:connections}b and is shown in Figure~\ref{fig:connections}b.

Finally, \tech{} supports the implementation of the receptive field using \textbf{Gaussian} connections, to
improve the performance of deep neural networks~\cite{luo2016understanding}.
Here, each neuron of a layer receives input only from neurons within a predefined region in its preceding layer. 
A receptive field is defined in Equation~\ref{eq:connections}c and is shown in Figure~\ref{fig:connections}c.

These basic connection modalities can be combined and extended to implement skip connections~\cite{tong2017image} and other types of connection found in modern SNNs. 

\subsubsection{Synaptic Polarity (Excitatory vs. Inhibitory)}
The \textbf{polarity parameter} \ineq{\beta_{ij}} is used to implement the inhibitory/excitatory effect of a synaptic connection on neuron activation. 
This enables a neural network to function as a universal approximator~\cite{wang2023networks}.
The polarity parameter is defined as:

\begin{equation}
    \label{eq:inh_exc}
    \footnotesize \beta_{ij} = \begin{cases}
        1 & \textbf{: excitatory effect}\\
        -1 & \textbf{: inhibitory effect}
    \end{cases}
\end{equation}

\subsection{Signed Neuronal Computations}
\tech{} uses fixed-point representations for all internal signals with configurable quantization and decimal precision.
We use the generalized \textbf{\texttt{Qn.q}} notation to represent a signed binary number with \ineq{n} integer bits and \ineq{q} fraction bits~\cite{yates2009fixed}, where \ineq{n} and \ineq{q} are the design parameters.
\tech{} uses the 2's compliment format to represent its internal signals, where the most significant bit (MSB) is used to encode the sign.
This signed representation along with signal computations allows the excitatory and inhibitory polarities of its synaptic weights.

\begin{figure}[h!]
    \centering    \centerline{\includegraphics[width=0.99\columnwidth]{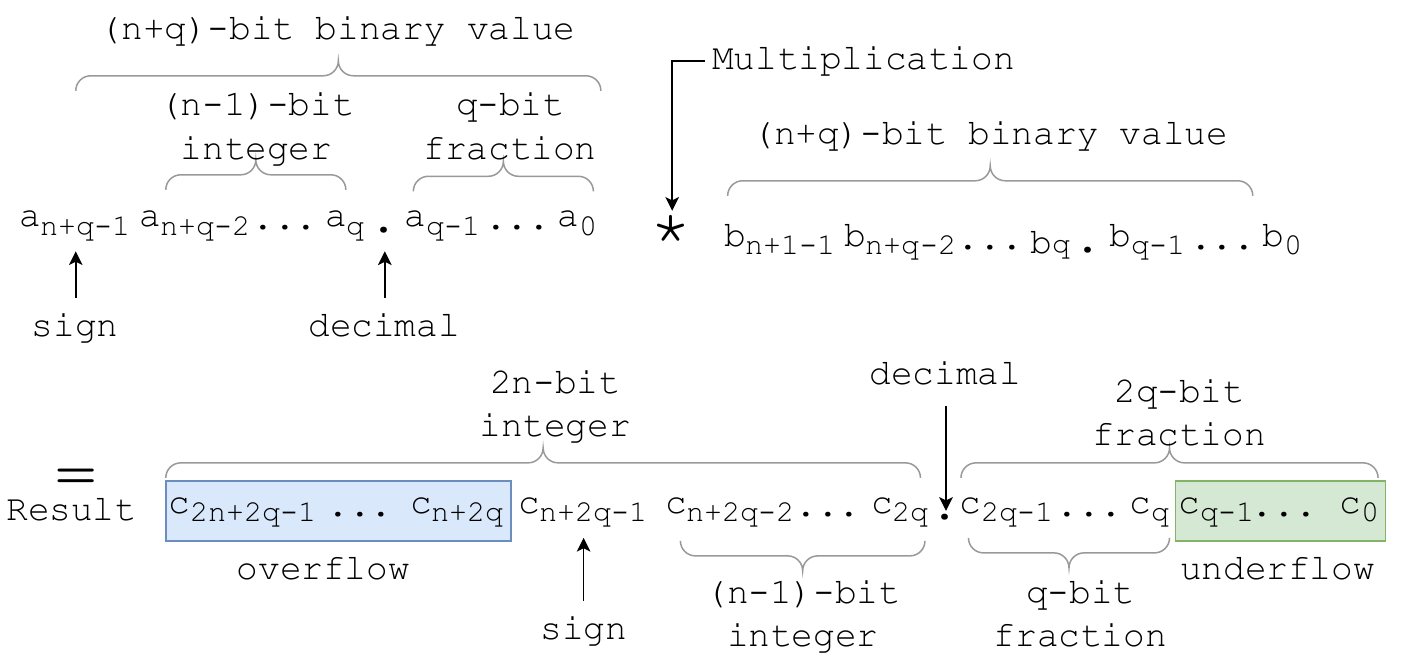}}
    \caption{Fixed-point multiplication of two $(n+q)$-bit numbers.}
    \label{fig:fp}
\end{figure}

Fixed-point addition and subtraction of two binary numbers follow the exact same rules as integer addition and subtraction, respectively.
However, fixed-point multiplication requires additional consideration, which we illustrate in Figure~\ref{fig:fp} using two \ineq{(n+q)}-bit binary numbers \textbf{a} and \textbf{b}.
The result of this multiplication uses \ineq{2n+2q} bits.
The final output is obtained by retaining \ineq{(n+q)} bits of the result.
The discarded most significant bits contribute to the overflow, whereas those from the least significant bits contribute to the underflow.

\subsection{Configuration Summary}
Table~\ref{tab:configs} summarizes the different configurations of \tech{} and their corresponding design enabler.

\begin{table}[h!]
    \renewcommand{\arraystretch}{1.0}
    \setlength{\tabcolsep}{2pt}
	\centering
	{\fontsize{6}{10}\selectfont
		\begin{tabular}{C{1.5cm} l c c}
			\hline
                 \textbf{Configuration} & \textbf{Parameters} & \textbf{Implementation} &  \textbf{Enabler}\\
                 \hline
                 \multirow{4}{*}{static}
                 & number of layers ($K$) & & \\
                 & number of neurons per layer ($N$) & HDL & application \\
                 & layer-to-layer connections ($\alpha,\beta$) & parameters & software\\
                 & precision and quantization ($n,q$) & &\\
                 \hline
                 \multirow{4}{*}{dynamic}
                 & growth rate & & application\\
                 & decay rate & control & software \&\\
                 & threshold voltage ($V_{th}$) & registers & system\\
                 & refractory period & & software\\
                \hline
	\end{tabular}}
        \vspace{5pt}
        \caption{Different configurations of \tech{}.}
	\label{tab:configs}
\end{table}


\section{Hardware-Software Interface}\label{sec:software}
\begin{figure}[h!]
    \centering
    \centerline{\includegraphics[width=0.99\columnwidth]{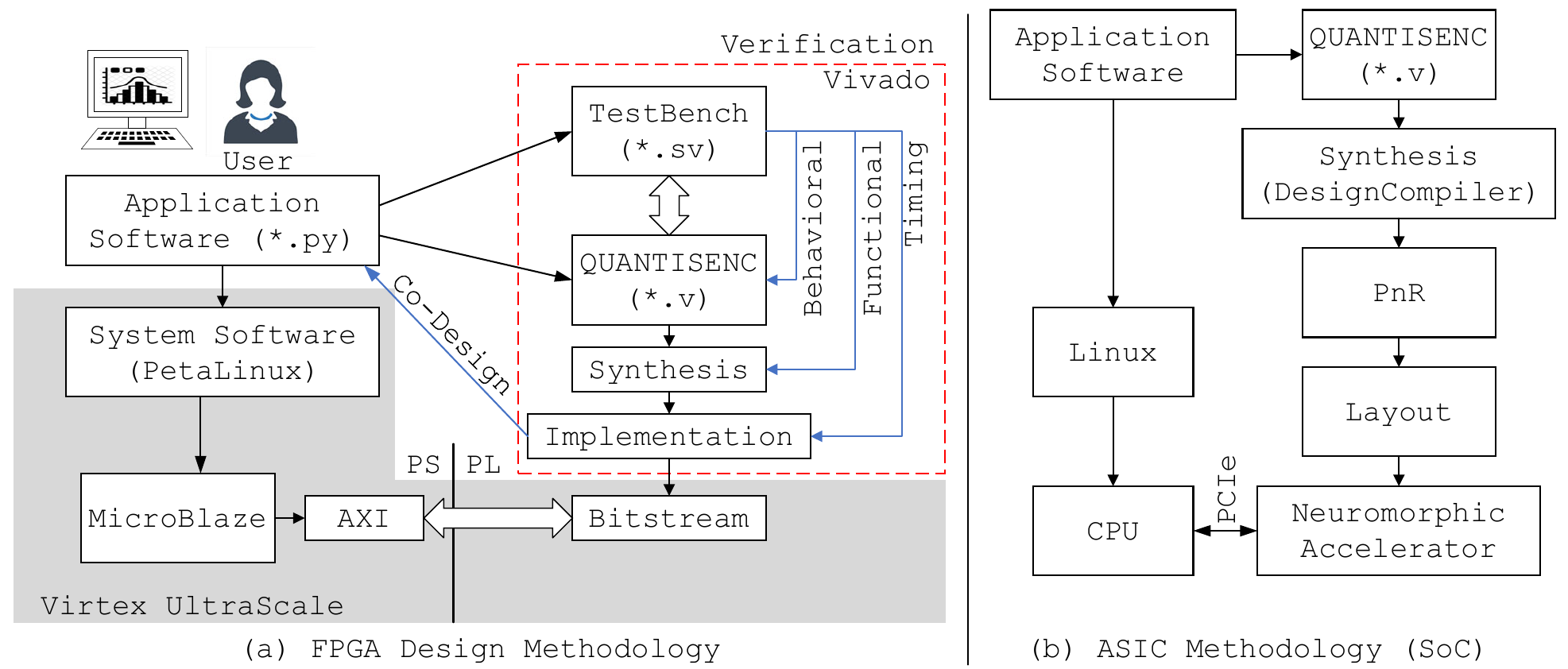}}
    \caption{Hardware-software interface of \tech{}.}
    \label{fig:flow}
\end{figure}

Figure~\ref{fig:flow}a shows the hardware-software interface of \tech{} designed for an FPGA-based computing system.
User code, which forms the application software, is written using PyTorch-based SNN constructs~\cite{eshraghian2023training,spikingjelly,spilger2023hxtorch}.
With moderate effort, \tech{} can be extended to support other SNN frameworks such as Brian~\cite{goodman2009brian} and CARLsim~\cite{ijcnn2022}.

In our system architecture, the application software runs on the processing system (PS) of the FPGA. 
It interfaces with the hardware implemented on the programmable logic (PL).
For AMD's Virtex Ultrascale evaluation board, the PS is composed of MicroBlaze with PetaLinux OS.
For AMD's Zynq MPSoC products, the MicroBlaze soft core can be replaced with ARM Cortex cores.
We designed open-source Python libraries 
to generate the configuration parameters of \tech{} in Verilog Hardware Description Language (HDL) and a testbench in SystemVerilog (sv) language.
The testbench is used to verify the design through (1) behavioral simulation (using the HDL), (2) functional simulation (using synthesized netlist), and (3) timing simulation (using implemented netlist) in AMD's Vivado tool suite.
\tech{}'s bitstream is implemented on the PL. 
The MicroBlaze software stack interfaces with the PL block to program synaptic weights, drive input, and visualize hardware results using the high-speed AMBA AXI interconnect.
We use timing simulations to extract the toggle rate of the nets, which is then used to calculate the average dynamic power consumption.
This power estimation framework can be integrated inside a loop (shown to the left) to do power/energy-aware co-design.

Finally, Figure~\ref{fig:flow}b shows an ASIC design flow using \tech{} targeted for accelerator-based designs.
The ASIC flow consists of front-end design steps that include synthesis and static timing analysis and back-end design steps that include placement, routing, and layout.
After fabrication and testing, the neuromorphic accelerator is connected to the host CPU inside a system-on-chip (SoC) using peripheral interconnect express (PCIe).
The hardware-software interface consists of the application software and the system software (e.g., Linux OS).
A key aspect of our design methodology is that the application software and the \tech{} hardware are the same for both the FPGA and ASIC flows.
Section~\ref{sec:asic} provides early results obtained using ASIC synthesis.

A key novelty of \tech{} is that synaptic memory is distributed among hardware layers. 
This distribution allows the layers to operate in parallel.
Although \tech{} performs layer-by-layer processing of spikes for any given input data stream (inherently a dataflow approach), we exploit pipelined parallelism in the design by overlapping the processing of streaming data via the system software~\cite{tecs2023}.
This is in addition to batch-level parallelism that is supported using \tech{} cores in a multi-/many-core setting.
\footnote{A batch-level parallelism is a type of parallel computing environment in a neuromorphic hardware where multiple batches of input data are processed concurrently on different processing elements~\cite{memocode2022}.}

\begin{figure}[h!]
    \centering
    \vspace{-5pt}
    \centerline{\includegraphics[width=0.99\columnwidth]{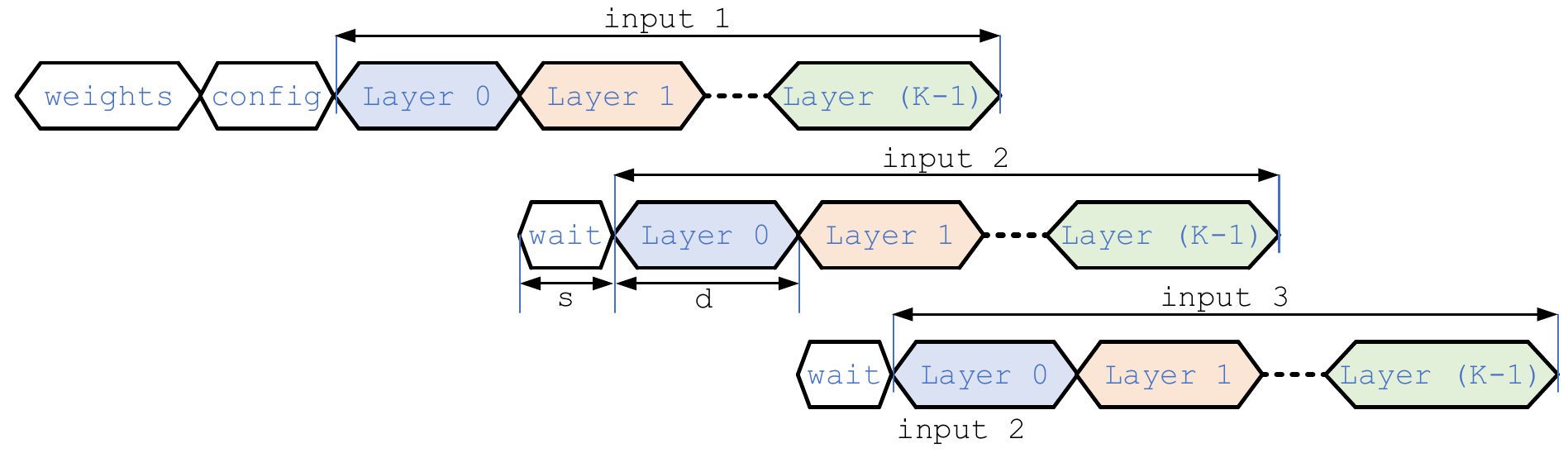}}
    \caption{Pipelined data processing using the hardware-software interface.}
    \label{fig:pipelining}
\end{figure}

We illustrate the processing of input data streams in Figure~\ref{fig:pipelining}.
Before execution begins, the synaptic weights and design parameters are loaded into the synaptic memory and control registers, respectively. 
Thereafter, data streams are scheduled one after another with a delay equal to the processing time of a layer (\ineq{d}) and a waiting time (\ineq{s}), as illustrated in the figure.
This waiting time ensures that the membrane potential of a neuron resets to its resting potential before it begins processing spike trains from the next input stream. In steady state, the maximum throughput obtained using this pipelined parallelism is \ineq{1 / (d + s)}.

\section{Distinction from State-of-the-Art}\label{sec:compare}
\begin{figure}[h!]
    \centering
    \centerline{\includegraphics[width=0.99\columnwidth]{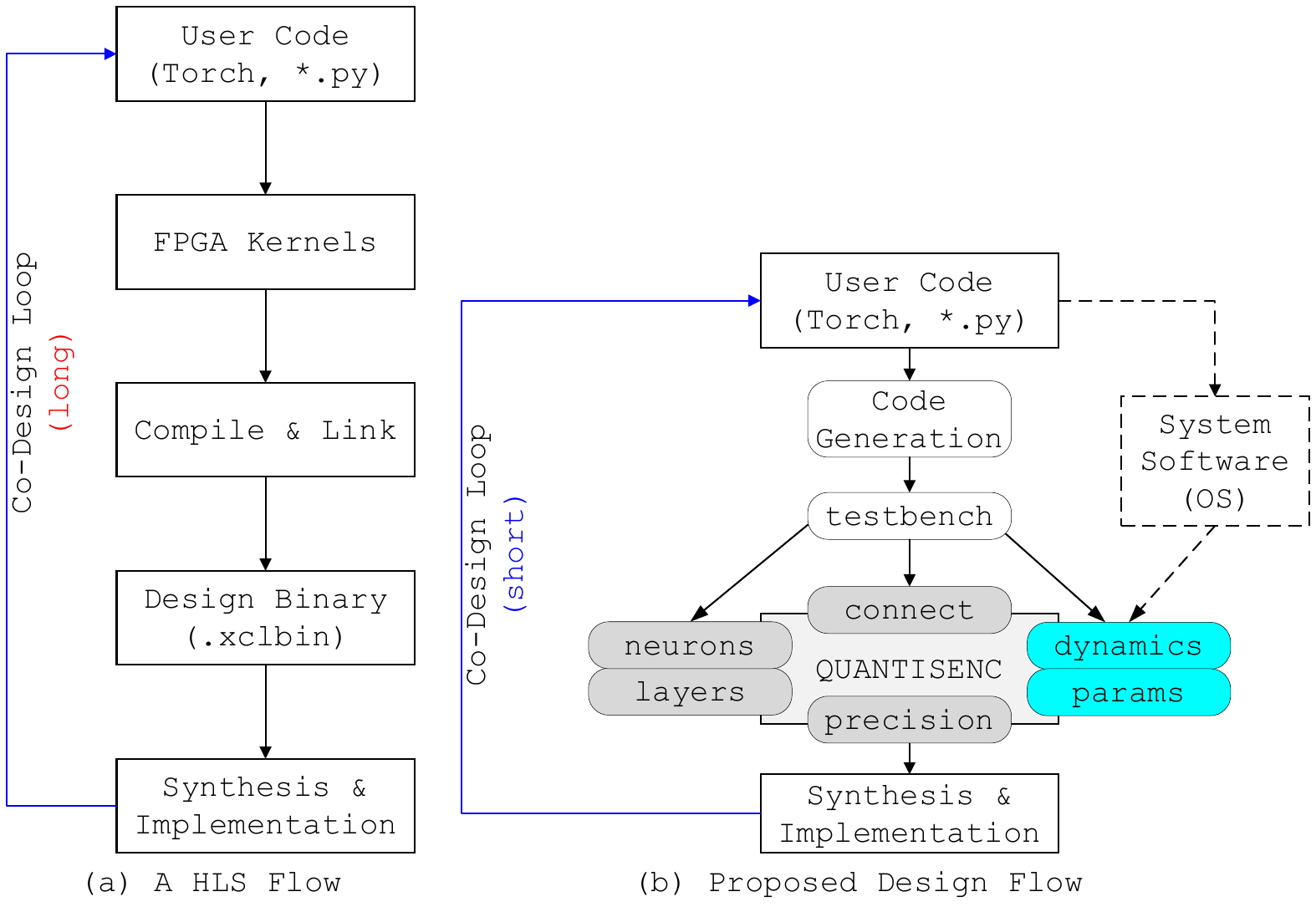}}
    \caption{(a) A High-Level Synthesis (HLS) flow within a co-design loop. (b) Our design methodology integrating \tech{} within a co-design loop.}
    \label{fig:compare}
\end{figure}

Although high-level synthesis (HLS) using commercial tools such as Vitis~\cite{kathail2020xilinx} and academic platforms such as Bambu~\cite{ferrandi2021bambu} can generate HDL starting from a software specification, there are four key limitations of neuromorphic HLS that we address using \tech{}. 
First, the HDL generated from HLS is not parameterizable. 
Therefore, any change in software description requires regenerating its HDL using the HLS flow, as illustrated in Figure~\ref{fig:compare}a. 
Large run times prohibit the use of HLS flow inside a co-design loop, where several hundred iterations may be necessary to achieve the required hardware and software performance for a target SNN model.
Second, most HLS tools do not support neuromorphic computations, and those that do support require a significant amount of engineering effort at the kernel level for compilation and mapping, in addition to the effort needed in hardware development to add new computations or extend existing ones, as shown in our recent work~\cite{iccad2021b}.
Third, most HLS tools work with dataflow graphs extracted from machine learning applications and perform static scheduling on them, exploiting only batch-level hardware parallelism. 
There remains a significant opportunity to improve performance, such as throughput, by taking advantage of other forms of parallelism that are present in neuromorphic hardware.
Finally, the HDL generated from HLS is not ASIC compatible~\cite{liu2022overgen,cong2022fpga,ahangari2023hls}.

Figure~\ref{fig:compare}b shows our design methodology.
\tech{} can be customized to match the exact specification of a user code (application software) simply by changing the design parameters.
Our fully automated design flow from model specification in software to hardware design and verification
is the core of our hardware-software co-design.
It does not require optimization at the kernel level, unlike an HLS.
This reduces the time to evaluate each design refinement, which reduces the overall co-design time.

Table~\ref{tab:compare} compares \tech{} with recent state-of-the-art designs in terms of relevant neuromorphic features. 
A comprehensive treatment of all existing designs is beyond the scope of this paper.
Here, we compare key techniques.

\begin{table*}[h!]
    \renewcommand{\arraystretch}{1.0}
    \setlength{\tabcolsep}{2pt}
	\centering
	{\fontsize{6}{10}\selectfont
		\begin{tabular}{c c C{1.2cm} C{1.0cm} C{2.0cm} C{1.0cm} C{1.0cm} C{1.5cm} C{2.5cm} C{2.0cm} c C{1.0cm}}
			\hline
                 & \textbf{Year} & \textbf{Operations} & \textbf{Neuron Dynamics} & \textbf{Reset Mechanism} & \textbf{Refractory Period} & \textbf{Synapse Model} & \textbf{Static Configuration} & \textbf{Dynamic Configuration} & \textbf{Programming Interface} & \textbf{Parallelism} & \textbf{Open Source}\\
                 \hline
                 
                 \cite{abdelsalam2018efficient} & 2018 & Floating Point (32~bits) & -- & -- & -- & Excitatory & Model Size & $\times$ & Custom C-based & $\times$ & $\times$\\
                 
                 \rowcolor{lavender} 
                 \cite{li2021fast} & 2021 & Floating Point (16~bits) & Nonlinear & Reset-to-Constant & $\surd$ & Excitatory & Model Size & $\times$ & Bare Metal & $\times$ & $\times$\\
                 
                 \cite{corradi2021gyro} & 2021 & -- & Linear & Reset-to-Zero & $\surd$ & Excitatory & Model Size, Quantization & $\times$ & Custom Python-based & $\times$ & $\times$\\
                 

                \rowcolor{lavender}
                \cite{carpegna2022spiker} & 2022 & Fixed Point (16~bits) & Nonlinear & Reset-to-Constant & -- & Excitatory & Model Size, Quantization & $\times$ & Bare Metal & $\times$ & $\times$\\
                
                \cite{liu2023low} & 2023 & Fixed Point (8~bits) & Linear & Reset-by-Subtraction & -- & Excitatory & Model Size, Quantization & $\times$ & Bare Metal & $\times$ & $\times$\\
                \hline
               \tech{} & 2023 & Fixed Point (\textbf{Qn.q}) & Nonlinear & Reset-to-Constant, Reset-to-Zero, Reset-by-Subtraction, Exponential Decay & -- & Excitatory \& Inhibitory & Model Size, Quantization, Connectivity & Decay Rate, Growth Rate, Threshold Voltage, Refractory Period & PyTorch & $\surd$ & $\surd$\\
                \hline
	\end{tabular}}
        \vspace{5pt}
        \caption{Comparing \tech{} against recent state-of-the-art designs in terms of key neuromorphic features.}
	\label{tab:compare}
\end{table*}


\section{Results}\label{sec:results}
\subsection{Evaluation Methodology}\label{sec:evaluation}
We evaluate \tech{} using Spiking MNIST~\cite{fatahi2016evt_mnist}, DVS Gesture~\cite{amir2017low}, and Spiking Heidelberg Digit (SHD)~\cite{cramer2020heidelberg} datasets on AMD's Virtex Ultrascale (primary), Virtex 7, and Zynq Ultrascale FPGA development boards.
Table~\ref{tab:datasets} summarizes these datasets and evaluation boards.
For each dataset, we report the number of classes, training, and test examples (in columns 1-4).
For each evaluation board, we report the technology and resources expressed in terms of Lookup Tables (LUTs), Flip-Flops (FFs), Block RAMs (BRAMs), and DSP Slices (in columns 5-7).

\begin{table}[h!]
    \renewcommand{\arraystretch}{1.0}
    \setlength{\tabcolsep}{1.2pt}
	\centering
	{\fontsize{6}{8}\selectfont
		\begin{tabular}{C{2.0cm} c C{1.0cm} C{1.0cm} | C{1.0cm} c C{1.0cm}}
			\hline 
                \textbf{Datasets} & \textbf{Classes} & \textbf{Training Examples} & \textbf{Test Examples} &
                \textbf{Evaluation Board} & \textbf{Technology} & \textbf{Resources}\\
                \hline
                Spiking MNIST~\cite{fatahi2016evt_mnist} & 10 & 60,000 & 100 & 
                Virtex UltraScale & 16nm FinFET & LUTs:~537,600 FFs:~1,075,200 BRAMs:~1728 DSPs:~768\\ 
                \hline
                DVS Gesture~\cite{amir2017low} & 11 & 1176 & 288 & 
                Virtex 7 & 28nm & LUTs:~303,600 FFs:~607,200 BRAMs:~1030 DSPs:~2800\\
                \hline
                Spiking Heidelberg Digit (SHD)~\cite{cramer2020heidelberg} & 20 & 8156 & 2264 & 
                Zynq UltraScale & 16nm FinFET & LUTs:~230,400 FF:~460,800 BRAM:~312 DSP:~1728\\ 
                \hline
	\end{tabular}}
        \vspace{5pt}
        \caption{Datasets and FPGA boards used to evaluate \tech{}.}
	\label{tab:datasets}
\end{table}

We use SNNTorch~\cite{eshraghian2023training} to train SNN models on a Lambda workstation, which has AMD Threadripper 3960X with 24 cores, 128 MB cache, 128 GB RAM and 2 RTX3090 GPUs.
The trained weights are programmed into the \tech{} hardware (in its synaptic memory) to perform classifications.

\begin{figure*}[h!]
    \centering
    \centerline{\includegraphics[width=1.99\columnwidth]{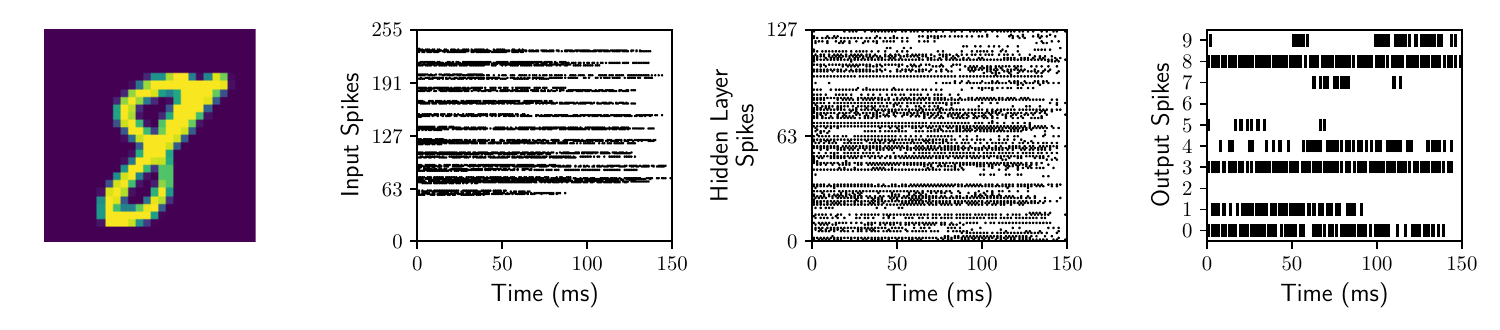}}
    \caption{A classification example with the spiking MNIST dataset using the handwritten digit 8. \tech{} is configured as a $256\times 128\times 10$ architecture.}
    \label{fig:example_8}
\end{figure*}

Figure~\ref{fig:example_8} shows a classification example using the spiking MNIST dataset for the handwritten digit 8.
The configuration of \tech{} is set to \ineq{(256\times 128\times 10)} with a quantization of 8 bits and a decimal precision of 3 bits. i.e., using the Q5.3 fixed-point representation.\footnote{We scale the original $28\times 28$ images to $16\times 16$ images to reduce model size.
This is the maximum scaling that can be performed on handwritten images from the dataset without affecting the classification accuracy.
}
This baseline configuration of \tech{} gives the best trade-off for the Spiking MNIST dataset in terms of application performance, e.g., accuracy of digit recognition (96.5\%), and hardware performance, e.g., area, power, latency, and throughput.

\begin{figure}[h!]
    \centering
    \centerline{\includegraphics[width=0.99\columnwidth]{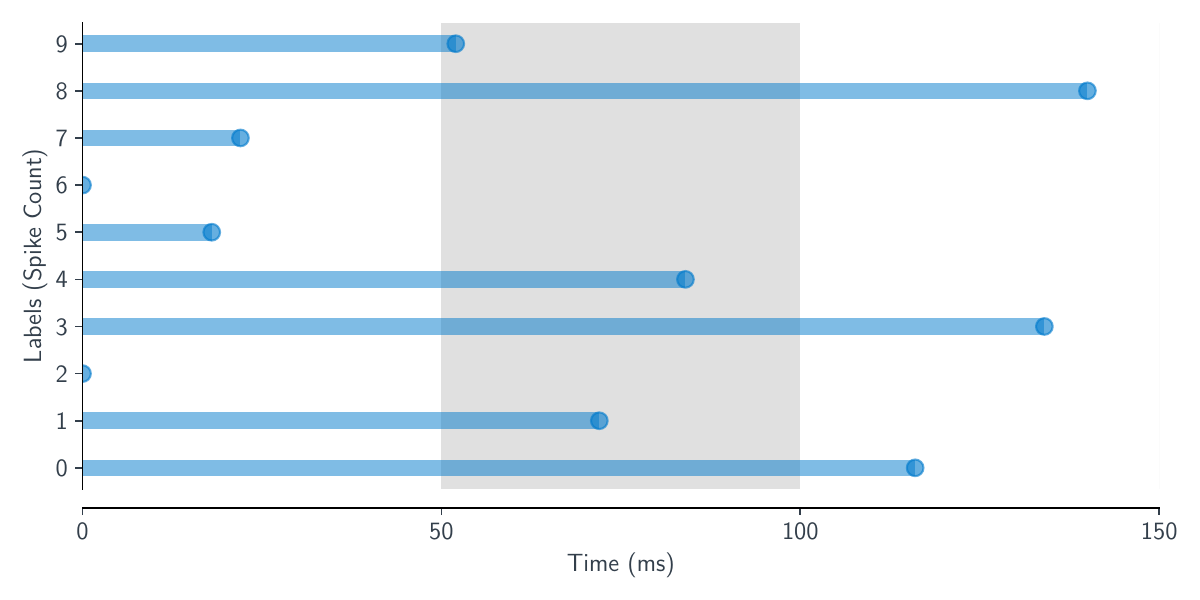}}
    \caption{Counting output spikes to decode \tech{} results.}
    \label{fig:animation}
\end{figure}

In Figure~\ref{fig:example_8}, the image is presented to \tech{} for a time duration of 150 ms.
This is a user-defined parameter and is controlled through the application software.
The figure shows the spikes generated by neurons of the three layers (input, hidden, and output).
We use a spike counter on the output layer to decode and visualize the classification result of \tech{}.
This is illustrated in Figure~\ref{fig:animation}.

We observe that output neuron 8 has the highest spike count, so the \tech{} result is correctly interpreted as 8 corresponding to the image.
However, neuron 3 has the second highest number of spikes, followed by neuron 0.
This is because of the structural similarity of the handwritten digit 8 to digits 3 and 0. 
When analyzing the confusion matrix, we observe that in most cases where a handwritten digit 8 is predicted incorrectly, it is predicted as a digit 3 or a digit 0.

\subsection{LIF Design}\label{sec:lif_results}
Table~\ref{tab:lif_results} reports the utilization of resources and the dynamic peak power of our LIF design for five different quantization and precision settings using a spike frequency of 100 MHz. 
We make the following three key observations.

\begin{table}[h!]
    \renewcommand{\arraystretch}{1.2}
    \setlength{\tabcolsep}{2 pt}
	\centering
	{\fontsize{7}{11}\selectfont
		\begin{tabular}{c C{2.5cm} | c c c c C{1.7cm}}
			\hline
                    & \textbf{Quantization \& Precision} & \textbf{LUTs} & \textbf{FFs} & \textbf{BRAMs} & \textbf{DSPs} & \textbf{Dynamic Peak Power (mW)}\\
                \hline
                    1 & binary & 14 & 11 & 0 & 0 & 3\\
                    2 & 4 bits (Q2.2) & 66 & 19 & 0 & 0 & 4\\
                    3 & 8 bits (Q5.3) & 245 & 35 & 0 & 0 & 6\\
                    4 & 16 bits (Q9.7) & 242 & 68 & 0 & 2 & 14\\
                    5 & 32 bits (Q17.15) & 856 & 132 & 0 & 8 & 27\\
                \hline
	\end{tabular}}
        \vspace{5pt}
        \caption{Resource utilization of the proposed LIF design for different settings of quantization and precision.}
	\label{tab:lif_results}
\end{table}


First, resource utilization (LUTs, FFs, etc.) increases with increasing quantization.
A 32-bit quantized LIF design uses 61x more LUTs and 12x more FFs compared to a 2-bit quantized design. 
This increase is due to the complexity of building an accurate LIF model in hardware.
Second, the increase in dynamic power, on the other hand, is less significant than resource utilization.
This is due to the extensive use of clock gating in the design, which minimizes dynamic power.
A 32-bit quantized LIF design consumes 9x more dynamic power than a 2-bit quantized design.
Third, DSP modules are utilized in the implementation for a quantization of 16 bits or higher.

\subsection{Neural Connections}\label{sec:connection_results}
Table~\ref{tab:lif_connections} reports the utilization of resources and dynamic power for three different connections supported in \tech{} -- one-to-one in row 1, Gaussian (convolution) in rows 2 \& 3, and all-to-all (fully connected) in rows 4, 5, \& 6.

\begin{table}[h!]
    \renewcommand{\arraystretch}{1.2}
    \setlength{\tabcolsep}{2 pt}
	\centering
	{\fontsize{7}{11}\selectfont
		\begin{tabular}{cc c | c c c c C{1.7cm}}
			\hline
                    &  & \textbf{Connections} & \textbf{LUTs} & \textbf{FFs} & \textbf{BRAMs} & \textbf{DSPs} & \textbf{Dynamic Peak Power (mW)}\\
                \hline
                    1 & one-to-one & $1$ & 296 & 56 & 0 & 0 & 12\\
                    \hline
                    2 & \multirow{2}{*}{convolution} & $3\times 3$ & 284 & 80 & 0.5 & 0 & 17\\
                    3 & & $5\times 5$ & 300 & 130 & 0.5 & 0 & 18\\
                    \hline
                    4 & \multirow{3}{*}{fully connected} & $128$ & 420 & 443 & 0.5 & 0 & 23\\
                    5 & & $256$ & 551 & 829 & 0.5 & 0 & 29\\
                    6 & & $512$ & 822 & 1599 & 0.5 & 0 & 48\\
                \hline
	\end{tabular}}
        \vspace{5pt}
        \caption{Resource utilization and peak dynamic power for different connection modalities in \tech{}.}
	\label{tab:lif_connections}
\end{table}

\begin{table*}[t!]
    \renewcommand{\arraystretch}{1.3}
    \setlength{\tabcolsep}{2 pt}
	\centering
	{\fontsize{7}{11}\selectfont
		\begin{tabular}{c c | c c | C{1.5cm} | C{1.5cm} c | C{1.5cm} c | C{1.5cm} c | C{1.2cm} c | C{1.5cm} c}
			\hline
                    &
                    \textbf{Configuration} & 
                    \textbf{Neurons} & \textbf{Synapses} &
                    \textbf{Quantization \& Precision} &
                    \textbf{LUTs (of~537,600)} & $\mathbf{\Delta}$ &
                    \textbf{FFs (of~1,075,200)} & $\mathbf{\Delta}$ &
                    \textbf{BRAMs (of~1728)} & $\mathbf{\Delta}$ &
                    \textbf{DSPs (of~768)} & $\mathbf{\Delta}$ &
                    \textbf{Dynamic Power (W)} & $\mathbf{\Delta}$ \\                    
                \hline
                
                \rowcolor{lavender}
                1 &
                $256\times 128\times 10$ & 
                394 & 34,048 & 
                Q5.3 & 
                8.97\% & -- & 
                0.98\% & -- & 
                3.99\% & -- & 
                0\% & -- & 
                0.623 & --\\
                \hline
                
                2 &
                $256\times 128\times 10$ & 
                394 & 34,048 & 
                Q9.7 & 
                9.38\% & 4.5\% & 
                1.39\% & 42.2\% & 
                3.99\% & 0\% & 
                35.93\% & -- & 
                0.738 & 18.5\%\\

                3 &
                $256\times 256\times 10$ & 
                522 & 68,096 & 
                Q5.3 & 
                17.44\% & 1.9$\times$ & 
                1.85\% & 1.9$\times$ & 
                7.69\% & 1.9$\times$ & 
                0\% & -- & 
                1.241 & 2.0$\times$\\

                4 &
                $256\times 256\times 256\times 10$ & 
                778 & 133,632 & 
                Q5.3 & 
                34.08\% & 3.8$\times$ & 
                3.55\% & 3.6$\times$ & 
                15.10\% & 3.8$\times$ & 
                0\% & -- & 
                2.172 & 3.5$\times$\\
                \hline
                



                \hline
	\end{tabular}}
        \vspace{5pt}
        \caption{Resource utilization and dynamic power of \tech{} for different SNN architectures.}
	\label{tab:resource}
\end{table*}

We make the following three observations.
First, a neuron with a one-to-one connection to another neuron uses 296 LUTs and 56 FFs for the quantization and precision of Q5.3.
The total dynamic power is 12 mW (after implementation).
The utilization of resources is comparable to that of a single neuron (see row 3 of Table~\ref{tab:lif_results}) with additional resources to implement synaptic weight between pre-synaptic and post-synaptic neurons.
Second, the Gaussian connection can be utilized to implement convolution filters.
We report the utilization and peak dynamic power of the $3\times 3$ and $5\times 5$ filters.
We observe that although there are more synaptic weights in a $3\times 3$ convolution filter than in a one-to-one connection, the former configuration uses fewer LUTs.
This is because it uses BRAMs to implement synaptic weights compared to the latter configuration, which uses LUTs to implement synaptic weight.
Finally, we report fully connected configurations with 128, 256, and 512 pre-synaptic connections per neuron.
We observe that with $2\times$ more connections (row 5 vs. 4), the number of LUTs and FFs increases by $1.3\times$ and $1.87\times$, respectively.
The peak dynamic power increases by only 6 mW.
On the other hand, with $4\times$ more connections (row 6 vs. 4), the number of LUTs and FFs increases by $1.95\times$ and $3.6\times$, respectively.
The dynamic peak power increases by $2\times$.

The results presented in Sections~\ref{sec:lif_results} \& \ref{sec:connection_results} clearly show the scalability of \tech{} in terms of its neuron configuration and connection modalities.
In the following, we provide an in-depth architectural analysis of \tech{} for the Spiking MNIST dataset using the baseline configuration of \ineq{(256\times 128\times 10)} with a quantization and precision of Q5.3 (Sections \ref{sec:resource}--\ref{sec:dynamic}).
We summarize such architectural analysis for the other two datasets in Section~\ref{sec:datasets}.

\subsection{Architectural Analysis for Spiking MNIST Dataset}\label{sec:resource}

Row 1 of Table~\ref{tab:resource} reports the utilization and dynamic power consumption of \tech{} for the baseline configuration of \ineq{256\times 128\times 10} and a quantization of 8 bits with a decimal precision of 3 bits, i.e., using Q5.3 fixed-point representation.
The design consists of 394 LIF neurons and 34,048 synapses.
It uses BRAMs to implement synaptic memory.
The utilization of resources for the Virtex UltraScale FPGA is as follows: 48,246 (= 8.97\%) LUTs, 10,550 (= 0.98\%) FFs, and 69 (= 3.99\%) BRAMs.
No DSP slices are used.
Dynamic power consumption is 623 mW using a spike frequency of 600 KHz.
This frequency gives the maximum performance per watt for the design (see Section~\ref{sec:throughput}).

Row 2 reports the utilization when the quantization is increased to 16 bits and the decimal precision to 7 bits, i.e., using Q9.7 fixed-point representation.
Increasing quantization and precision results in an increase of 4.5\% in LUTs and 42.2\% in FFs compared to the baseline configuration in row~1.
The number of BRAMs remains the same, but the design uses 276 (= 35.93\%) DSPs.
Dynamic power increases by 18.5\%.

Rows 3 \& 4 show the utilization and dynamic power for two additional design configurations.
The configuration in row~3 has 522 neurons (32.5\% higher than baseline) and 68,096 synapses (2\ineq{\times} higher than baseline).
For this configuration, the design requires 1.9$\times$ more LUTs, 1.9$\times$ more FFs, and 1.9$\times$ more BRAMs than the baseline.
The dynamic power is 2$\times$ higher.
On the other hand, the configuration in row 4 has 778 neurons (2$\times$ higher than baseline) and 133,632 synapses (4\ineq{\times} higher than baseline).
For this configuration, the design requires 3.8$\times$ more LUTs, 3.6$\times$ more FFs, and 3.8$\times$ more BRAMs than the baseline.
The dynamic power is 3.5$\times$ higher.

These results show the scalability of \tech{} with the number of neurons and synapses.
One way to utilize this scalability result would be to quickly evaluate the utilization of FPGA resources for a specific configuration of \tech{}, without having to synthesize the design, which often takes a considerable amount of time.
This is useful when conducting design space exploration, where multiple iterations may be necessary before making a final design choice.

\subsection{Comparison to State-of-the-Art}\label{sec:sota}
Table~\ref{tab:sota} compares 
resource utilization, power, and accuracy for single neurons and the SNN architecture for the Spiking MNIST dataset against latest state-of-the-art designs.
We make the following four key observations.

First, for a single neuron (columns 2, 3, \& 4), the Euler approximation in the proposed implementation uses more LUTs and FFs than~\cite{guo2021toward} and \cite{ye2022implementation}.
This is because, in the proposed design, a neuron's refractory period, reset mechanism, growth rate, decay rate, and threshold voltage can all be configured at run-time using a combination of the application and system software.
The extra logic to implement such configurability increases the resource utilization, but at the same time provides flexibility in exploring the trade-off between accuracy and power consumption even after a design is implemented on silicon.
Neuron implementation in~\cite{guo2021toward} \& \cite{ye2022implementation} can only be configured statically. 
Once these designs are implemented on silicon, they do not offer run-time configurability.
Second, the power consumption of the proposed design is lower than~\cite{guo2021toward}
due to the extensive use of clock gating.
Specifically, we gate the clock in the design when there is no input spike. 

Third, for a complete SNN architecture, the proposed design uses fewer neurons and synapses to achieve an accuracy comparable to that of the state-of-the-art~\cite{abdelsalam2018efficient,he2021low}.
Fourth, the highest accuracy (reported in the literature) is obtained using~\cite{abdelsalam2018efficient}.
However, this design consumes 3.4 W of dynamic power.
Compared to this design, our design achieves slightly lower accuracy (96.5\% vs. 98.4\%) while consuming only 623 mW of dynamic power.
Finally, the most optimized hardware design is proposed in~\cite{he2021low}.
The resource utilization is significantly lower because it uses a high-level synthesis approach to optimize the design, including removing extra design logic that is required to configure the design at run-time.
In addition to the lack of configurability and ASIC portability, this design consumes 1.03 W of dynamic power and achieves an accuracy of 93\%.
Compared to this, the proposed design consumes less power and achieves higher accuracy.

We conclude that the proposed design achieves accuracy comparable to state-of-the-art designs while using fewer neurons and synapses. It also consumes less dynamic power.

\begin{table}[h!]
    \renewcommand{\arraystretch}{1.0}
    \setlength{\tabcolsep}{2pt}
	\centering
	{\fontsize{6}{8}\selectfont
		\begin{tabular}{c c c c | C{1.3cm} C{1.3cm} c}
			\hline 
                    & \multicolumn{3}{c|}{\textbf{A Single Neuron}} & \multicolumn{3}{c}{\textbf{SNN Architecture}}\\
                    & 
                    \textbf{Euler~\cite{guo2021toward}} & \textbf{Euler~\cite{ye2022implementation}} & 
                    \textbf{Ours} & \textbf{Best Accuracy~\cite{abdelsalam2018efficient}} & \textbf{Best Hardware~\cite{he2021low}} &  \textbf{Ours}\\
                \hline
                Configuration &  -- & -- & -- & 784-1024-10 & 784-2048-10 & 256-128-10\\
                Neurons & -- &  -- & -- & 1818 & 2932 & 394\\
                Synapses & -- & -- & -- & 813,056 & 1,810,432 & 34,048\\
                \hline
                LUTs & 95 & 76 & 108 & 78,679 & 16,813 & 40,965\\
                FFs & 85 & 20 & 23 & 16,864 & 7,559 & 7,095\\
                BRAMs & 0 & 0 & 0 & 174 & 129 & 69\\
                \hline
                Power (W) & 0.25 & NR & 0.05 & 3.4 & 1.03 & 0.623\\
                Accuracy & -- & -- & -- & 98.4\% & 93.0\% & 96.5\%\\
                \hline
	\end{tabular}}
        \vspace{5pt}
        \caption{comparison to state-of-the-art.}
	\label{tab:sota}
\end{table}

\subsection{Software vs. Hardware Performance}\label{sec:software_vs_hardware}
\begin{figure}[h!]
    \centering
    \centerline{\includegraphics[width=1.00\columnwidth]{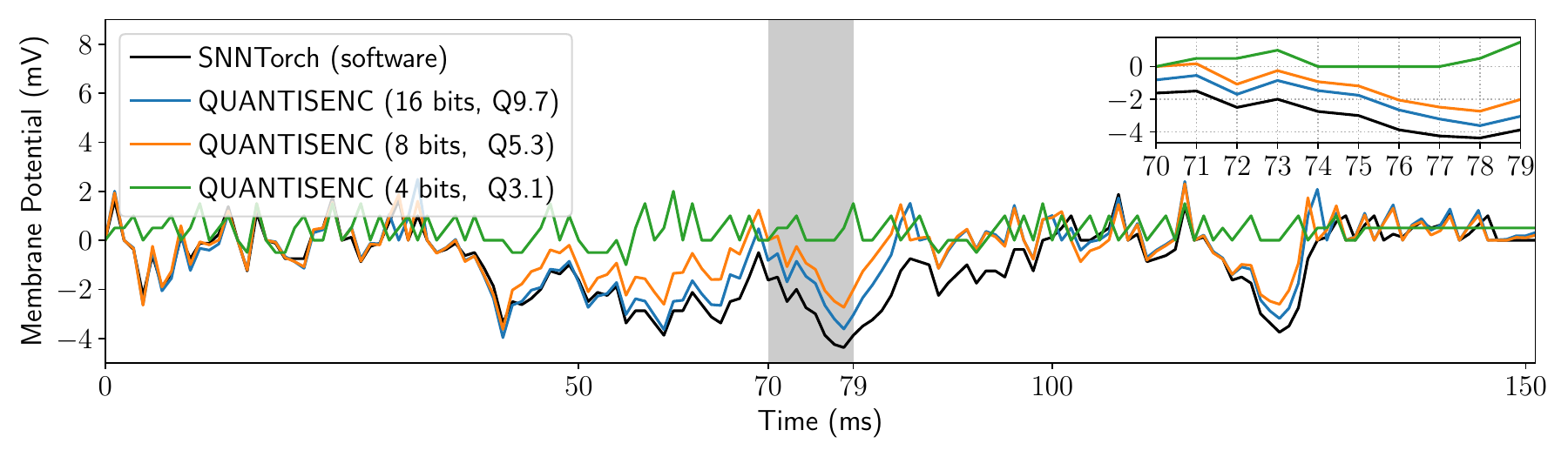}}
    \caption{Impact of quantization on neuron membrane potential.}
    \label{fig:vmem}
\end{figure}

Figure~\ref{fig:vmem} plots the membrane potential of a neuron in the hidden layer of \tech{} with the baseline configuration of \ineq{(256\times 128\times 10)} for the handwritten digit 8 image from the spiking MNIST dataset. 
We evaluate three quantization and precision settings: Q9.7 (16-bit quantization with 7-bit decimal precision), Q5.3 (8-bit quantization with 3-bit decimal precision), and Q3.1 (4-bit quantization with 1-bit decimal precision).
For reference, we also include the membrane potential obtained using SNNTorch (software), which uses a double-precision floating-point representation.

For easier comparison of the results, the subplot inside the plot zooms in on the time scale between 70 and 80 ms.
We observe that the membrane potential using the Q9.7 fixed-point representation is close to that using the double precision floating-point representation.
The average root mean square error (RMSE) of all 394 neurons for 100 randomly selected test examples from the spiking MNIST dataset is 0.25 mV.
The average RMSE increases to 0.43 mV using the Q5.3 representation.
Finally, the RMSE using the Q3.1 representation is 2.12 mV.

Table~\ref{tab:software_vs_hardware} reports the accuracy comparison between SNNTorch (software) and \tech{} (hardware) for the Spiking MNIST dataset. 
We compare results for three different quantization and precision settings: Q9.7, Q5.3, and Q3.1.
We make the following three key observations.

\begin{table}[h!]
    \renewcommand{\arraystretch}{1.0}
    \setlength{\tabcolsep}{2pt}
	\centering
	{\fontsize{6}{8}\selectfont
		\begin{tabular}{c c C{1.5cm} C{1.5cm} C{1.5cm}}
			\hline 
                \multirow{2}{*}{\textbf{Dataset}} & \textbf{SNNTorch} & \multicolumn{3}{c}{\textbf{Average Test Set Hardware Accuracy}}\\
                & \textbf{Accuracy} & \textbf{Q9.7} & \textbf{Q5.3} & \textbf{Q3.1}\\
                \hline
                Spiking MNIST & 97.8\% & 97.1\% & 96.5\% & 88.3\%\\
                \hline
	\end{tabular}}
        \vspace{5pt}
        \caption{Accuracy with different quantization and precision settings.}
	\label{tab:software_vs_hardware}
\end{table}

First, the accuracy with a quantization of 16 bits and a decimal precision of 7 bits (Q9.7) is comparable to 
SNNTorch software.
This is because the membrane potential value obtained in hardware using this setting is close to that in software (see Figure~\ref{fig:vmem})
due to our detailed LIF design (see Figure~\ref{fig:lif}).
Second, the accuracy reduces with a reduction in quantization and decimal precision.
The accuracy using 8-bit and 4-bit quantization is 96.5\% and 88.3\%, compared to 97.1\% using 16-bit quantization and 97.8\% using SNNTorch software.

\subsection{Design Throughput and Throughput Per Watt}\label{sec:throughput}
\begin{figure}[h!]
    \centering
    \centerline{\includegraphics[width=1.00\columnwidth]{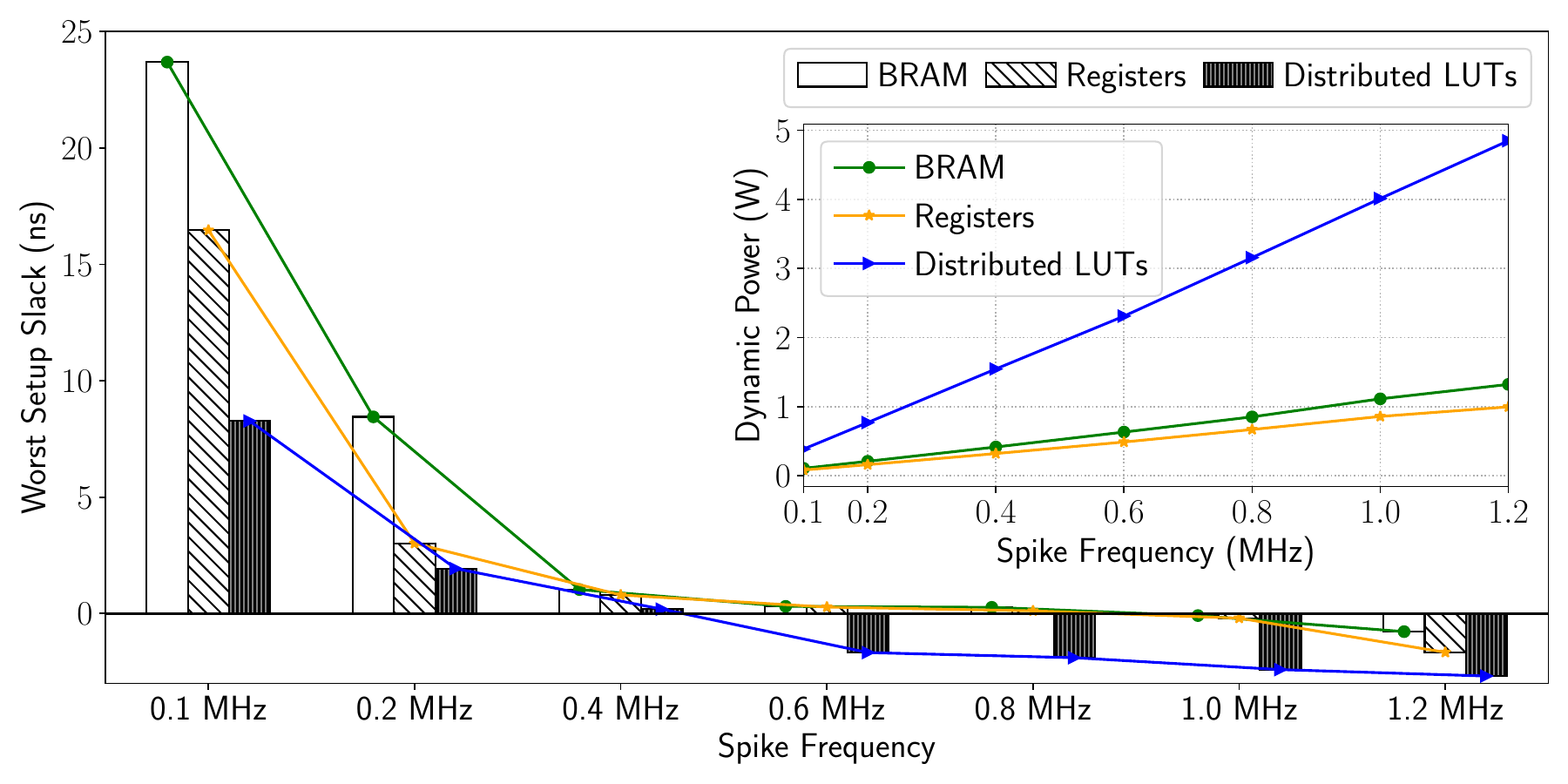}}
    \caption{Maximum frequency and power trade-off.}
    \label{fig:freq}
\end{figure}

Figure~\ref{fig:freq} plots the worst setup slack (in ns) of \tech{} as we increase the spike frequency from 100 KHz to 1.2 MHz using the baseline configuration of \ineq{(256\times 128\times 10)} and Q5.3 quantization and decimal precision.
Setup slack is defined as the difference between the required time and the arrival time of the data at an endpoint (typically a register).
During static timing analysis (STA), a negative setup slack indicates timing violations.
The peak frequency is one that results in the least positive setup slack.
Figure~\ref{fig:freq} reports the setup slack for three synaptic memory implementations: BRAM, registers, and distributed LUTs.
The subplot reports the dynamic power of \tech{} for these memory settings.
We make the following three key observations.

First, the setup slack is positive for a spike frequency of 100, 200, and 400 KHz for all three memory implementations.
As we increase the frequency to 600 KHz, there are multiple timing violations for register-based synaptic memory. 
The peak frequency for this implementation is 500 KHz.
Second, the setup slack for distributed LUT-based implementation is 61\% lower than for BRAM-based implementation, which means that the latter supports a higher peak frequency.
Our results show that the peak spike frequencies for these two implementations are 850 and 925 KHz, respectively.
Finally, the distributed LUT implementation has the least dynamic power for all spike frequencies.
It is 23\%  and 79\% lower than the BRAM and register-based implementations, respectively.

In addition to the peak frequency, we also evaluate \tech{} based on its real-time performance, which is measured as the number of images inferred per second, and throughput per watt, which is measured as the number of fixed-point operations performed per watt.
These are defined as
\begin{equation}
    \footnotesize \text{Real-time Performance} = \frac{1}{\text{exposure time} + N_\text{reset}/f} \text{ (see Sec.~\ref{sec:software})}
\end{equation}
where exposure time is the time interval for which each image is exposed to the SNN model for inference, \ineq{N_\text{reset}} is the number of clock cycles needed to reset the membrane potential, and \ineq{f} is the spike frequency.
The value \ineq{N_\text{reset}} depends on the membrane time constant (\ineq{\tau}).
Our empirical studies show that \ineq{N_\text{reset} = 4} clock cycles at \ineq{f = 1 \text{ KHz}} for \ineq{\tau = 5} ms.
This results in a real-time performance of 41.67 frames per second (fps) for an exposure time of 20 ms.
This is the performance obtained by exploiting pipelined parallelism in \tech{}.
In previous works such as~\cite{corradi2021gyro}, such parallelism has not been exploited.
Therefore, the real-time performance is \ineq{\frac{1}{\text{exposure time} + (K\times L) / f}}, where \ineq{K} is the number of layers and \ineq{L} is the latency (in the number of clock cycles) of each layer.
For the baseline design \ineq{(256\times 128\times 10)} with three layers, the maximum performance obtained using \cite{corradi2021gyro} is 31.25 fps.
We conclude that \tech{} improves the real-time performance by 33. 3\% by exploiting pipelined parallelism.

The throughput per watt is computed as the ratio of the total number of fixed-point operations per second to power consumption.
Due to pipeline execution in \tech{}, all synaptic accumulations in all layers occur in parallel, where each accumulation involves a single fixed-point operation.
In addition, all neuron computations also take place in parallel, working on different inputs.
Therefore, the total number of fixed-point operations in one second is
\begin{equation}
    \label{eq:fpop}
    \footnotesize \text{Fixed-Point Ops} = \left(N_\text{synapse} + N_\text{ops}\times N_\text{neurons}\right)\times f
\end{equation}
where \ineq{f} is the design frequency and \ineq{N_\text{ops}} is the number of fixed-point operations per neuron.

\begin{figure}[h!]
    \centering
    \centerline{\includegraphics[width=1.00\columnwidth]{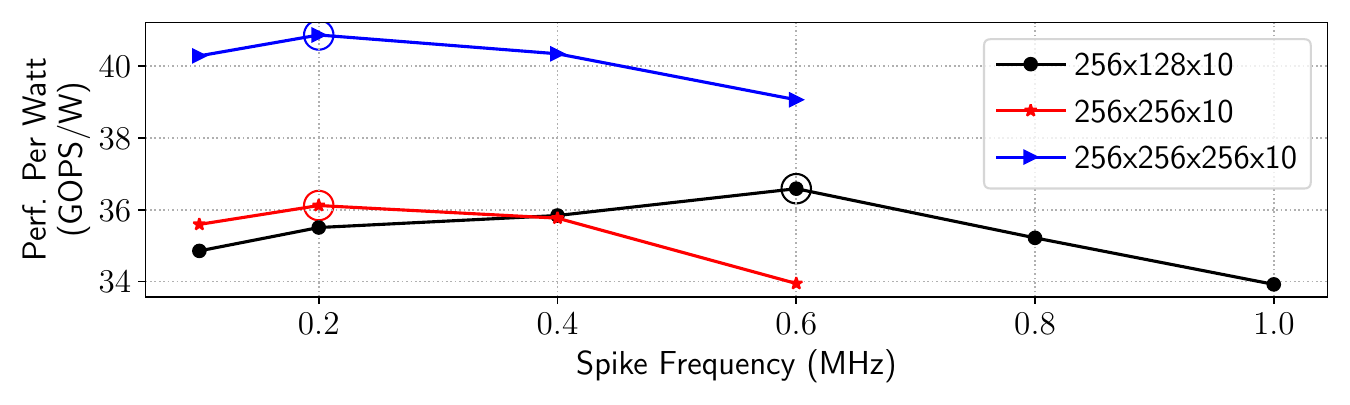}}
    \caption{Performance per watt for different SNN architectures using BRAMs.}
    \label{fig:gops}
\end{figure}

Figure~\ref{fig:gops} plots the performance per watt for the three designs evaluated in Table~\ref{tab:resource} as the spike frequency increases from 100 KHz to 1 MHz.
All these designs use BRAM for synaptic implementation.
We make the following observations.

First, the performance per watt increases with an increase in frequency until it reaches the maximum.
This is because at lower frequencies the increase in performance (see Equation~\ref{eq:fpop}) exceeds the increase in dynamic power, resulting in an overall increase in performance per watt for all evaluated designs.
With a further increase in frequency, the dynamic power starts to dominate the performance-power ratio, resulting in a reduction in performance per watt.
The maximum performance per watt for each design configuration is indicated with a circle in Figure~\ref{fig:gops}, which occurs at a frequency much lower than the highest frequency supported by the respective design.

\subsection{Static Configuration of \tech{}}\label{sec:settings}

Table~\ref{tab:static} reports the largest configuration of \tech{} that can be implemented on three evaluated FPGA platforms.
We report the configuration for a wide \tech{} design using a single hidden layer and a deep \tech{} design using multiple hidden layers.
We also report the power consumption of these configurations.
Our software-defined hardware design methodology allows us to easily explore different design configurations for a target FPGA platform.

\begin{table}[h!]
    \renewcommand{\arraystretch}{1.2}
    \setlength{\tabcolsep}{2pt}
	\centering
	{\fontsize{6}{8}\selectfont
		\begin{tabular}{c | C{2.0cm} c | C{2.0cm} c}
			\hline 
                    \multirow{2}{*}{\textbf{Platform}} & 
                    \multicolumn{2}{| c }{\textbf{Wide Design (Single Hidden Layer) }} &
                    \multicolumn{2}{| c}{\textbf{Deep Design (Multiple Hidden Layers)}}\\
                    & \textbf{Configuration} & \textbf{Power} & \textbf{Configuration} & \textbf{Power}\\
                \hline
                    Virtex UltraScale & 256-1470-10 & 9.557 W & 256-28(64)-10 & 6.371\\
                    Virtex 7 & 256-704-10 & 5.818 W & 256-20(64)-10 & 4.833\\
                    Zynq UltraScale & 256-640-10 & 3.349 W & 256-12(64)-10 & 1.854\\
                
                \hline
	\end{tabular}}
        \vspace{5pt}
        \caption{\tech{} configuration on three FPGA platforms.}
	\label{tab:static}
\end{table}

We observe that the Virtex UltraScale platform has more resources (LUTs, FFs, etc.) than the other two evaluated platforms (see Table~\ref{tab:datasets}).
Therefore, the largest configuration supported on this platform is also the highest for both wide and deep designs. 
To implement such a large design, the power consumption is also significantly higher.


\subsection{Dynamic Configuration of \tech{}}\label{sec:dynamic}

Table~\ref{tab:dynamic} reports the average spikes per neuron, dataset accuracy, and dynamic power for four R \& C settings (keeping the time constant $\tau$ = 5 ms), three reset mechanisms (default \ding{182}, reset-by-subtraction \ding{183} and reset-to-zero \ding{184}), and two refractory periods (of 0 and 5 cycles). 

\begin{table}[h!]
    \renewcommand{\arraystretch}{1.0}
    \setlength{\tabcolsep}{2.2pt}
	\centering
	{\fontsize{6}{8}\selectfont
		\begin{tabular}{r | C{0.8cm} C{0.8cm} C{0.8cm} C{0.8cm} | c c c | c c}
			\hline 
                    & 
                    \multicolumn{4}{|c}{\textbf{R, C Settings}}
                    & \multicolumn{3}{|c}{\textbf{Reset}}
                    & \multicolumn{2}{|c}{\textbf{Ref.}}\\
                    & 
                    R=500M$\Omega$ C=10pF &
                    R=100M$\Omega$ C=50pF &
                    R=50M$\Omega$ C=100pF &
                    R=10M$\Omega$ C=500pF &
                    \ding{182} & \ding{183} & \ding{184} &
                    0 & 5\\
                \hline
                    Avg. Spikes per Neuron & 26 & 19 & 7 & 0 & 45 & 26 & 22 & 26 & 20\\
                    Dataset Accuracy (\%) & 96.5 & 94.4 & 67.8 & -- & 92.7 & 96.5 & 96.5 & 96.5 & 95.8\\
                    Dynamic Power (mW) & 663 & 541 & 449 & -- & 1,087 & 663 & 625 & 663 & 580\\
                \hline
	\end{tabular}}
        \vspace{5pt}
        \caption{Impact of dynamic settings of \tech{}.}
	\label{tab:dynamic}
\end{table}

We make the following three observations.
First, the growth rate of the membrane potential of a neuron is high for a large R (500 M\ineq{\Omega}) and a small C (10 pF), resulting in an average of 26 spikes per neuron.
This is the same setting that was used in SNNTorch to train the model for the spiking MNIST dataset.
The accuracy is 96.5\%.
Reducing R to 100 M\ineq{\Omega} and increasing C to 50 pF results in a reduction in the number of spikes per neuron. 
Accuracy is reduced only marginally.
However, a reduction in the number of spikes results in a significant reduction in dynamic power (18.4\%).
With a further reduction in R (and increase in C), the number of spikes per image decreases.
Although the dynamic power is reduced by 33\%, the accuracy is much lower. 
With a small R (10M\ineq{\Omega}) and a large C (500 pF), the membrane potential does not exceed the threshold.
As a result, no spikes are generated.

Second, the baseline reset setting is reset-by-subtraction (column 7). 
Using the default reset mechanism increases the number of spikes per neuron, which reduces the accuracy and increases the dynamic power.
Reset-to-zero results in fewer spikes per neuron. 
The accuracy remains the same as the baseline configuration.
However, the dynamic power is lower.

Finally, the number of spikes per neuron decreases with increasing refractory period.
Using our design methodology, all these parameters can be configured at run-time to explore the trade-off between performance and power.

\subsection{Dataset Results}\label{sec:datasets}

Table~\ref{tab:dataset_summary} summarizes the results for the three evaluated datasets.
We show a thorough evaluation for the Spiking MNIST dataset in Sections \ref{sec:resource}--\ref{sec:dynamic}. 
These results are summarized in row 1 of Table~\ref{tab:dataset_summary}.

\begin{table}[h!]
    \renewcommand{\arraystretch}{1.2}
    \setlength{\tabcolsep}{1.2pt}
	\centering
	{\fontsize{6}{8}\selectfont
		\begin{tabular}{c C{1.0cm} c C{0.8cm} C{0.8cm} C{0.8cm} c C{1.0cm} C{1.4cm}}
			\hline 
                   & & \multirow{2}{*}{\textbf{Configuration}} & \multicolumn{3}{c}{\textbf{Resource Utilization}} & \multirow{2}{*}{\textbf{Accuracy}} & \textbf{Dynamic Peak Power} & \textbf{Peak Performance per Watt}\\
                    \cline{4-6}
                    & &  & LUTs & FFs & BRAMs &  &  \textbf{(W)} & \textbf{(GOPS/W)}\\ 
                \hline
                   1. & Spiking MNIST~\cite{fatahi2016evt_mnist} & 256-128-10 & 9\% & 1\% & 4\% & 96.5\% & 0.623 & 36.6\\
                   2. & DVS Gesture~\cite{amir2017low} & 400-300-300-11 & 60\% & 15\% & 18\% & 85.07\% & 1.827 & 24.45\\
                   3. & SHD~\cite{cramer2020heidelberg} & 700-256-256-20 & 65\% & 20\% & 24\% & 87.8\% & 1.629 & 16.09\\  
                   
                \hline 
	\end{tabular}}
        \vspace{5pt}
        \caption{Design summary for different datasets.}
	\label{tab:dataset_summary}
\end{table}

Row 2 summarizes our design exploration for the DVS Gesture dataset.
The design requires a configuration of \ineq{(400\times 300\times 300\times 11)}, which uses 60\% LUTs, 15\% FFs, and 18\% BRAMs on VirtexUltraScale. 
The accuracy obtained is 85.07\% compared to an SNNTorch accuracy of 87.1\%~\cite{he2020comparing}.
The implemented design has a peak dynamic power of 1.827 W and a performance per watt of 24.45.

Row 3 summarizes our design exploration for the SHD dataset.
The design requires a configuration of \ineq{(700\times 256\times 256\times 20)}, which uses 65\% LUTs, 20\% FFs, and 24\% BRAMs on VirtexUltraScale. 
The accuracy obtained is 87.8\% compared to the SNNTorch accuracy of 90.3\%~\cite{hammouamri2023learning}.
The implemented design has a peak dynamic power of 1.629 W and a performance per watt of 16.09.

\subsection{Early ASIC Synthesis Results}\label{sec:asic}

Table~\ref{tab:asic} reports the results of the early ASIC synthesis of the proposed LIF design using the Synopsys Design Compiler for a spike frequency of 100 MHz.
The proposed design uses 1,574 nets, 944 combinatorial cells, 35 sequential cells (Flip-Flops), and 309 buffers \& inverters, occupying a total area of 2894 $\mu m^2$.
The design consumes 101.7 $\mu$ W of power, divided into 23.2 $\mu$ W of switching (dynamic) power and 78.5 $\mu$W of leakage power.
Our future work will explore other stages of ASIC design as illustrated in Figure~\ref{fig:flow}(c).

\begin{table}[h!]
    \renewcommand{\arraystretch}{1.2}
    \setlength{\tabcolsep}{2pt}
	\centering
	{\fontsize{6}{8}\selectfont
		\begin{tabular}{c c C{0.8cm} C{0.8cm} c c C{1.0cm} C{1.0cm} C{1.0cm}}
			\hline 
                    \textbf{Technology} & \textbf{Nets} & \textbf{Comb. Cells} & \textbf{Seq. Cells} & \textbf{Buf/Inv} & \textbf{Area} & \textbf{Switching Power} & \textbf{Leakage Power} & \textbf{Total Power}\\
                \hline
                32nm & 1574 & 944 & 35 & 309 & 2894 $\mu m^2$ & 23.2 $\mu$W & 78.5 $\mu$W & 101.7 $\mu$W\\
                \hline
	\end{tabular}}
        \vspace{5pt}
        \caption{Early ASIC synthesis results of a Q5.3 LIF neuron.}
	\label{tab:asic}
\end{table}

\section{Conclusion} \label{sec: conclusion}
We propose \tech{}, a fully configurable open source neuromorphic core architecture designed to advance research in neuromorphic computing.
\tech{} distinguishes itself through its hierarchical design methodology, which allows system designers to configure key parameters such as the number of layers, neurons per layer, layer-to-layer connectivity, neuron dynamics, quantization, and precision through the software, both at the design time and at run-time. 
The proposed framework allows to define and train SNN models using PyTorch-based simulators and assess their hardware performance through FPGA prototyping and ASIC design. 
\tech{}'s performance has been evaluated using datasets like Spiking MNIST, DVS Gesture, and Spiking Heidelberg Digit (SHD) on AMD’s Virtex 7, Virtex Ultrascale, and Zynq Ultrascale FPGA development boards, demonstrating superior resource utilization, power efficiency, latency, and throughput compared to state-of-the-art designs.

\section*{Acknowledgments}
This material is based upon work supported by Accenture LLP, the U.S. Department of Energy under Award Number DE-SC0022014 and by the National Science Foundation under Grant
Nos. OAC-2209745, CCF-1942697, \& CNS-2008167.

\bibliographystyle{IEEEtran}
\bibliography{external,disco}

\begin{thebibliography}{10}
\providecommand{\url}[1]{#1}
\csname url@samestyle\endcsname
\providecommand{\newblock}{\relax}
\providecommand{\bibinfo}[2]{#2}
\providecommand{\BIBentrySTDinterwordspacing}{\spaceskip=0pt\relax}
\providecommand{\BIBentryALTinterwordstretchfactor}{4}
\providecommand{\BIBentryALTinterwordspacing}{\spaceskip=\fontdimen2\font plus
\BIBentryALTinterwordstretchfactor\fontdimen3\font minus \fontdimen4\font\relax}
\providecommand{\BIBforeignlanguage}[2]{{%
\expandafter\ifx\csname l@#1\endcsname\relax
\typeout{** WARNING: IEEEtran.bst: No hyphenation pattern has been}%
\typeout{** loaded for the language `#1'. Using the pattern for}%
\typeout{** the default language instead.}%
\else
\language=\csname l@#1\endcsname
\fi
#2}}
\providecommand{\BIBdecl}{\relax}
\BIBdecl

\bibitem{mead1990neuromorphic}
C.~Mead, ``Neuromorphic electronic systems,'' \emph{Proceedings of the IEEE}, vol.~78, no.~10, pp. 1629--1636, 1990.

\bibitem{joubert2012hardware}
A.~Joubert, B.~Belhadj, O.~Temam, and R.~H{\'e}liot, ``{Hardware spiking neurons design: Analog or digital?}'' in \emph{International Joint Conference on Neural Networks (IJCNN)}.\hskip 1em plus 0.5em minus 0.4em\relax IEEE, 2012.

\bibitem{maass1997networks}
W.~Maass, ``Networks of spiking neurons: The third generation of neural network models,'' \emph{Neural Networks}, vol.~10, no.~9, pp. 1659--1671, 1997.

\bibitem{indiveri2003low}
G.~Indiveri, ``A low-power adaptive integrate-and-fire neuron circuit,'' in \emph{International Symposium on Circuits and Systems (ISCAS)}, 2003.

\bibitem{ward2022beyond}
M.~Ward and O.~Rhodes, ``Beyond {LIF} neurons on neuromorphic hardware,'' \emph{Frontiers in Neuroscience}, vol.~16, p. 881598, 2022.

\bibitem{cessac2008dynamics}
B.~Cessac and T.~Vi{\'e}ville, ``On dynamics of integrate-and-fire neural networks with conductance based synapses,'' \emph{Frontiers in Computational Neuroscience}, vol.~2, p. 228, 2008.

\bibitem{fatahi2016evt_mnist}
M.~Fatahi, M.~Ahmadi, M.~Shahsavari, A.~Ahmadi, and P.~Devienne, ``evt\_mnist: A spike based version of traditional mnist,'' \emph{arXiv}, 2016.

\bibitem{amir2017low}
A.~Amir, B.~Taba, D.~Berg, T.~Melano, J.~McKinstry, C.~Di~Nolfo, T.~Nayak, A.~Andreopoulos, G.~Garreau, M.~Mendoza \emph{et~al.}, ``A low power, fully event-based gesture recognition system,'' in \emph{Conference on Computer Vision and Pattern Recognition (CVPR)}, 2017.

\bibitem{cramer2020heidelberg}
B.~Cramer, Y.~Stradmann \emph{et~al.}, ``{The Heidelberg spiking data sets for the systematic evaluation of spiking neural networks},'' \emph{IEEE Transactions on Neural Networks and Learning Systems (TNNLS)}, 2020.

\bibitem{gungor2022optimizing}
M.~Gungor, K.~Huang, S.~Ioannidis, and M.~Leeser, ``Optimizing designs using several types of memories on modern {FPGAs},'' in \emph{High Performance Extreme Computing (HPEC)}, 2022.

\bibitem{luo2016understanding}
W.~Luo, Y.~Li, R.~Urtasun, and R.~Zemel, ``Understanding the effective receptive field in deep convolutional neural networks,'' \emph{Advances in Neural Information Processing Systems (NeurIPS)}, 2016.

\bibitem{tong2017image}
T.~Tong, G.~Li, X.~Liu, and Q.~Gao, ``Image super-resolution using dense skip connections,'' in \emph{Intl. Conf. on Computer Vision (ICCV)}, 2017.

\bibitem{wang2023networks}
Q.~Wang, M.~A. Powell, A.~Geisa, E.~Bridgeford, C.~E. Priebe, and J.~T. Vogelstein, ``Why do networks have inhibitory/negative connections?'' in \emph{International Conference on Computer Vision (ICCV)}, 2023.

\bibitem{yates2009fixed}
R.~Yates, ``Fixed-point arithmetic: An introduction,'' \emph{Digital Signal Labs}, vol.~81, no.~83, p. 198, 2009.

\bibitem{eshraghian2023training}
J.~K. Eshraghian, M.~Ward, E.~O. Neftci, X.~Wang, G.~Lenz, G.~Dwivedi, M.~Bennamoun, D.~S. Jeong, and W.~D. Lu, ``Training spiking neural networks using lessons from deep learning,'' \emph{Proc. of the IEEE}, 2023.

\bibitem{spikingjelly}
W.~Fang, Y.~Chen, J.~Ding, Z.~Yu, T.~Masquelier, D.~Chen, L.~Huang, H.~Zhou, G.~Li, and Y.~Tian, ``{SpikingJelly: An open-source machine learning infrastructure platform for spike-based intelligence},'' \emph{Science Advances}, vol.~9, no.~40, p. eadi1480, 2023.

\bibitem{spilger2023hxtorch}
P.~Spilger, E.~Arnold, L.~Blessing, C.~Mauch, C.~Pehle, E.~M{\"u}ller, and J.~Schemmel, ``hxtorch. snn: Machine-learning-inspired spiking neural network modeling on brainscales-2,'' in \emph{Neuro-Inspired Computational Elements Conference}, 2023.

\bibitem{goodman2009brian}
D.~F. Goodman and R.~Brette, ``The brian simulator,'' \emph{Frontiers in Neuroscience}, vol.~3, p. 643, 2009.

\bibitem{ijcnn2022}
L.~Niedermeier, K.~Chen, J.~Xing, A.~Das, J.~Kopsick, E.~Scott, N.~Sutton, K.~Weber, N.~Dutt, and J.~L. Krichmar, ``{CARLsim 6: an open source library for large-scale, biologically detailed spiking neural network simulation},'' in \emph{International Joint Conference on Neural Networks (IJCNN)}, 2022.

\bibitem{tecs2023}
A.~Das, ``A design flow for scheduling spiking deep convolutional neural networks on heterogeneous neuromorphic system-on-chip,'' \emph{ACM Transactions on Embedded Computing Systems (TECS)}, 2023.

\bibitem{memocode2022}
------, ``Real-time scheduling of machine learning operations on heterogeneous neuromorphic {SoC},'' in \emph{International Conference on Formal Methods and Models for System Design (MEMOCODE)}, 2022.

\bibitem{kathail2020xilinx}
V.~Kathail, ``Xilinx vitis unified software platform,'' in \emph{International Symposium on Field Programmable Gate Arrays (FPGA)}, 2020.

\bibitem{ferrandi2021bambu}
F.~Ferrandi, V.~G. Castellana, S.~Curzel, P.~Fezzardi, M.~Fiorito, M.~Lattuada, M.~Minutoli, C.~Pilato, and A.~Tumeo, ``Bambu: An open-source research framework for the high-level synthesis of complex applications,'' in \emph{Design Automation Conference (DAC)}, 2021.

\bibitem{iccad2021b}
S.~Curzel, N.~B. Agostini, S.~Song, I.~Dagli, A.~Limaye, C.~Tan, M.~Minutoli, V.~G. Castellana \emph{et~al.}, ``Automated generation of integrated digital and spiking neuromorphic machine learning accelerators,'' in \emph{International Conference on Computer-Aided Design (ICCAD)}, 2021.

\bibitem{liu2022overgen}
S.~Liu, J.~Weng, D.~Kupsh, A.~Sohrabizadeh, Z.~Wang, L.~Guo, J.~Liu, M.~Zhulin, R.~Mani, L.~Zhang \emph{et~al.}, ``{OverGen: Improving FPGA usability through domain-specific overlay generation},'' in \emph{International Symposium on Microarchitecture (MICRO)}, 2022.

\bibitem{cong2022fpga}
J.~Cong, J.~Lau, G.~Liu, S.~Neuendorffer, P.~Pan, K.~Vissers, and Z.~Zhang, ``{FPGA HLS today: Successes, challenges, and opportunities},'' \emph{ACM Trans, on Reconfigurable Technology and Systems (TRETS)}, 2022.

\bibitem{ahangari2023hls}
H.~Ahangari, M.~M. {\"O}zdal, and {\"O}.~{\"O}zt{\"u}rk, ``{HLS}-based high-throughput and work-efficient synthesizable graph processing template pipeline,'' \emph{ACM Trans. on Embedded Computing Systems (TECS)}, vol.~22, 2023.

\bibitem{abdelsalam2018efficient}
A.~M. Abdelsalam, F.~Boulet, G.~Demers, J.~P. Langlois, and F.~Cheriet, ``An efficient {FPGA}-based overlay inference architecture for fully connected {DNNs},'' in \emph{International Conference on Reconfigurable Computing and FPGAs (ReConFig)}, 2018.

\bibitem{li2021fast}
S.~Li, Z.~Zhang, R.~Mao, J.~Xiao, L.~Chang, and J.~Zhou, ``A fast and energy-efficient {SNN} processor with adaptive clock/event-driven computation scheme and online learning,'' \emph{IEEE Transactions on Circuits and Systems I: Regular Papers}, vol.~68, no.~4, pp. 1543--1552, 2021.

\bibitem{corradi2021gyro}
F.~Corradi, G.~Adriaans, and S.~Stuijk, ``Gyro: A digital spiking neural network architecture for multi-sensory data analytics,'' in \emph{Drone Systems Engineering and Rapid Simulation and Performance Evaluation: Methods and Tools}, 2021.

\bibitem{carpegna2022spiker}
A.~Carpegna, A.~Savino, and S.~Di~Carlo, ``{Spiker: An FPGA-optimized hardware accelerator for spiking neural networks},'' in \emph{IEEE Computer Society Annual Symposium on VLSI (ISVLSI)}, 2022.

\bibitem{liu2023low}
H.~Liu, Y.~Chen, Z.~Zeng, M.~Zhang, and H.~Qu, ``A low power and low latency {FPGA}-based spiking neural network accelerator,'' in \emph{International Joint Conference on Neural Networks (IJCNN)}, 2023.

\bibitem{guo2021toward}
W.~Guo, H.~E. Yant{\i}r, M.~E. Fouda, A.~M. Eltawil, and K.~N. Salama, ``Toward the optimal design and {FPGA} implementation of spiking neural networks,'' \emph{IEEE Transactions on Neural Networks and Learning Systems}, vol.~33, no.~8, pp. 3988--4002, 2021.

\bibitem{ye2022implementation}
W.~Ye, Y.~Chen, and Y.~Liu, ``The implementation and optimization of neuromorphic hardware for supporting spiking neural networks with mlp and cnn topologies,'' \emph{IEEE Transactions on Computer-Aided Design of Integrated Circuits and Systems}, vol.~42, no.~2, pp. 448--461, 2022.

\bibitem{he2021low}
Z.~He, C.~Shi, T.~Wang, Y.~Wang, M.~Tian, X.~Zhou, P.~Li, L.~Liu, N.~Wu, and G.~Luo, ``A low-cost {FPGA} implementation of spiking extreme learning machine with on-chip reward-modulated {STDP} learning,'' \emph{IEEE Transactions on Circuits and Systems II: Express Briefs}, vol.~69, 2021.

\bibitem{he2020comparing}
W.~He, Y.~Wu, L.~Deng, G.~Li, H.~Wang, Y.~Tian, W.~Ding, W.~Wang, and Y.~Xie, ``{Comparing SNNs and RNNs on neuromorphic vision datasets: Similarities and differences},'' \emph{Neural Networks}, 2020.

\bibitem{hammouamri2023learning}
I.~Hammouamri, I.~Khalfaoui-Hassani, and T.~Masquelier, ``Learning delays in spiking neural networks using dilated convolutions with learnable spacings,'' \emph{arXiv preprint arXiv:2306.17670}, 2023.

\end{thebibliography}


 




\begin{IEEEbiography}[{\includegraphics[width=1in,height=1.25in,clip,keepaspectratio]{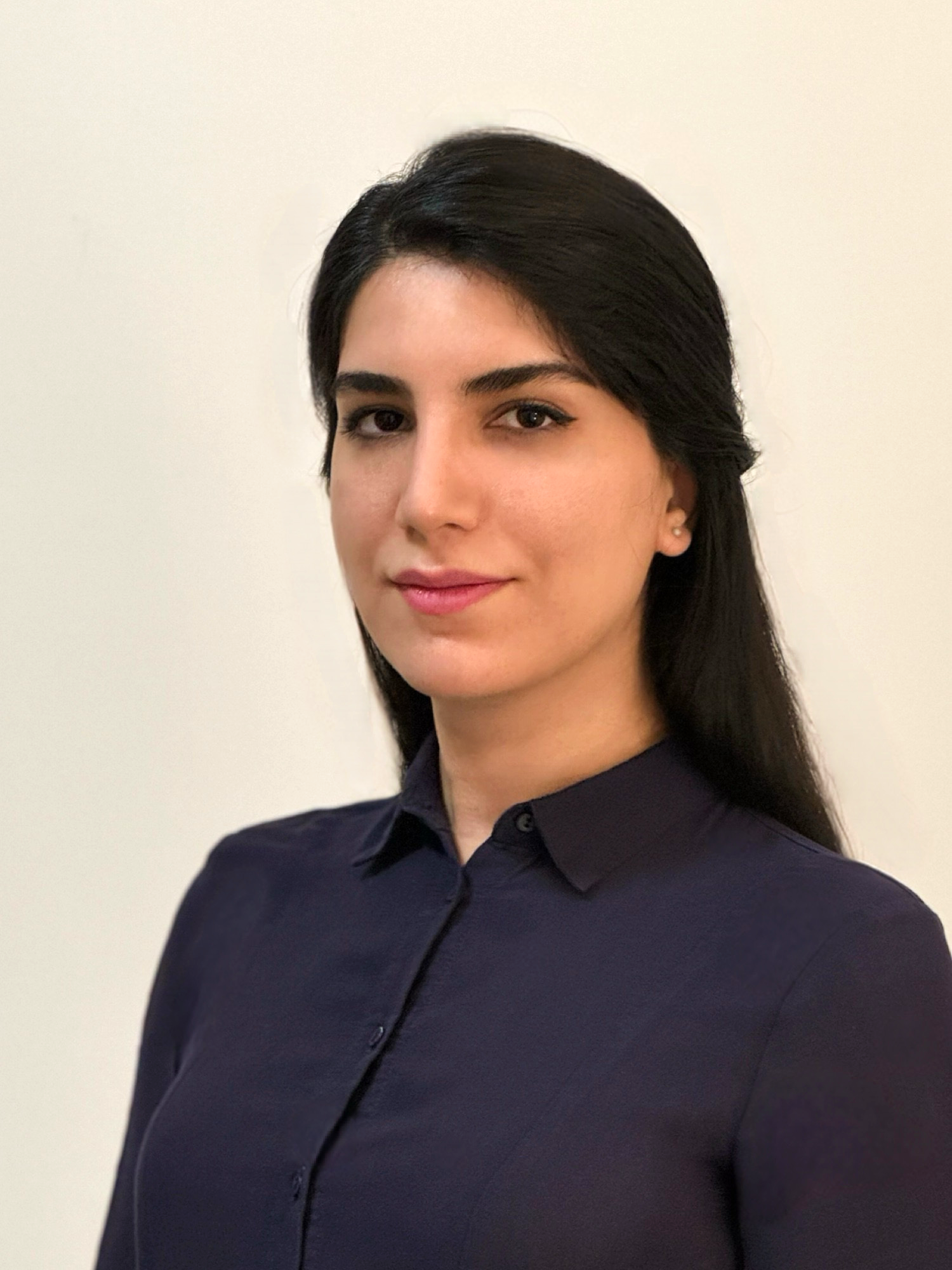}}]{Shadi Matinizadeh}
received her master's degree in electrical engineering from California State University, Northridge, in 2023. She is currently working toward the PhD degree under the supervision of Dr.Anup Das from the Department of Electrical and Computer Engineering, Drexel University, Philadelphia. Her research interests include neuromorphic computing and spiking neural networks.
\end{IEEEbiography}

\begin{IEEEbiography}[{\includegraphics[width=1in,height=1.25in,clip,keepaspectratio]{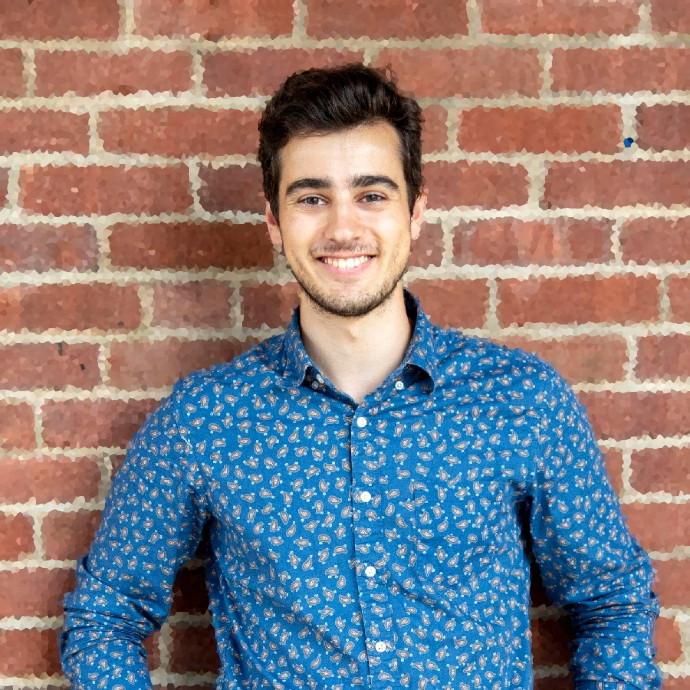}}]{Noah~Pacik-Nelson}
is a Research Scientist at Accenture Labs. He focuses on neuromorphic computing, IoT devices, and next-gen 3D printing. He received his BSE in Computer Science from the University of Connecticut, Storrs CT in 2020.
\end{IEEEbiography}

\begin{IEEEbiography}[{\includegraphics[width=1in,height=1.25in,clip,keepaspectratio]{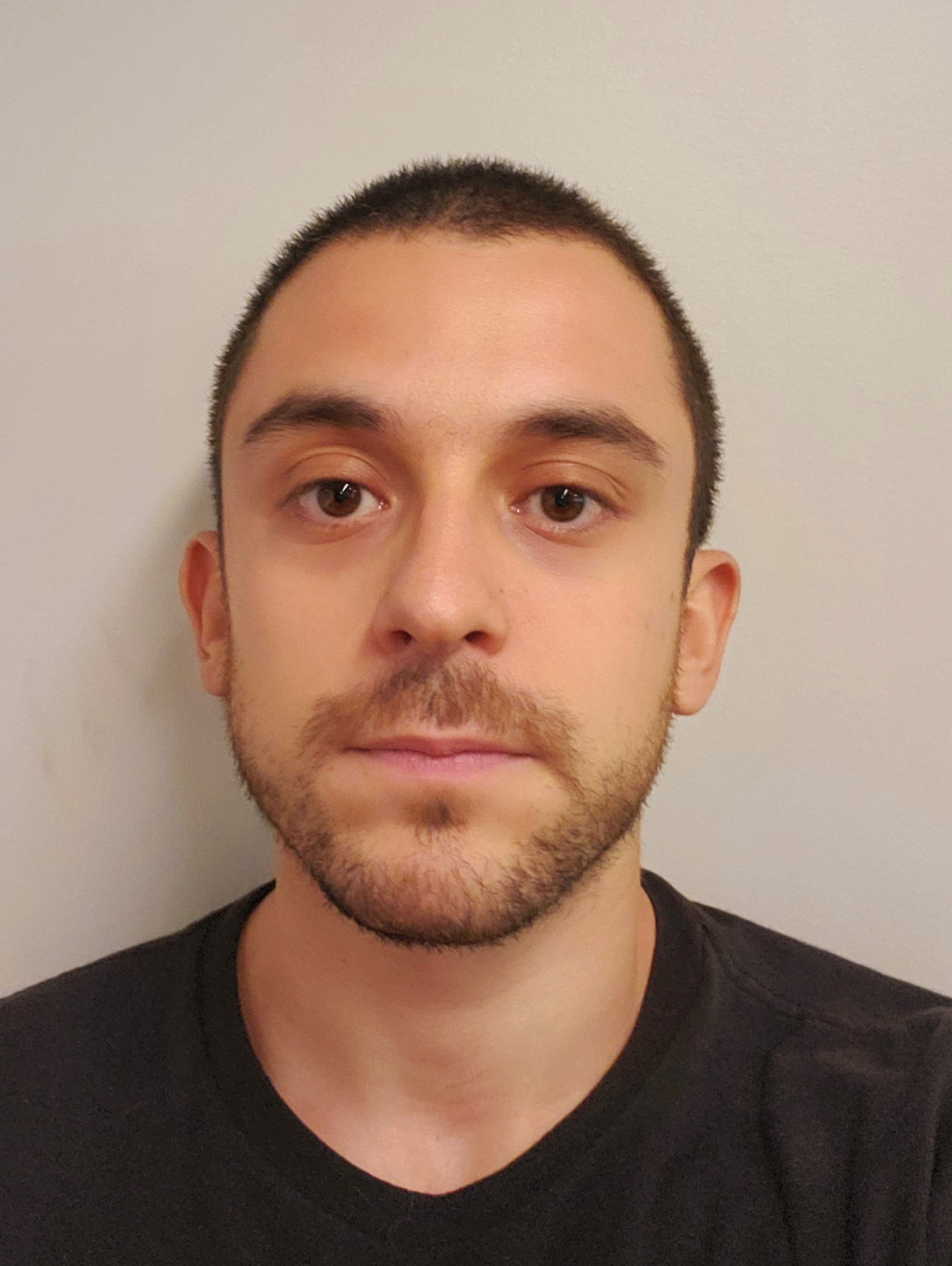}}]{Ioannis~Polykretis}
is a R\&D Associate Principal at Accenture Labs in San Francisco, CA, USA. His research interests include low-power edge processing, brain-inspired computing, neuromorphic engineering, and robotics. He received his B.S. and M.S. degrees in Electrical and Computer Engineering from the National Technical University of Athens in 2016. He received his Ph.D. in Computer Science from Rutgers University, New Brunswick, NJ, in 2023.
\end{IEEEbiography}

\begin{IEEEbiography}[{\includegraphics[width=1in,height=1.25in,clip,keepaspectratio]{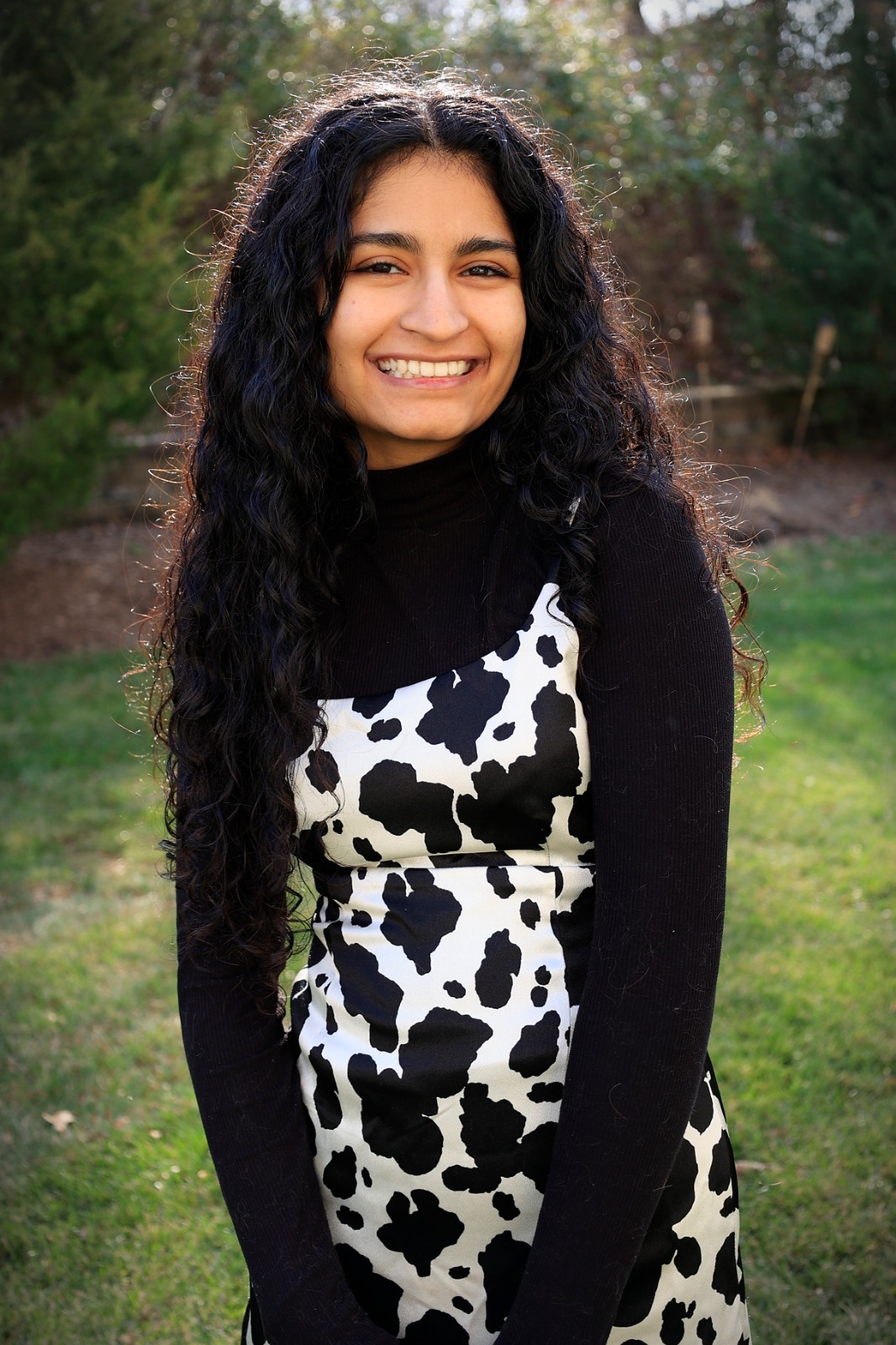}}]{Krupa Tishbi}
earned two degrees, a B.S. in Electrical Engineering and a B.S. in Physics, from The College of New Jersey in 2022. Her undergraduate research involved designing control architectures on an FPAA board, as well as digital video and image processing with neural networks. She is now working towards her Ph.D. in Electrical Engineering at Drexel University. Her graduate research involves the ASIC design of spiking neural networks.
\end{IEEEbiography}

\begin{IEEEbiography}[{\includegraphics[width=1in,height=1.25in,clip,keepaspectratio]{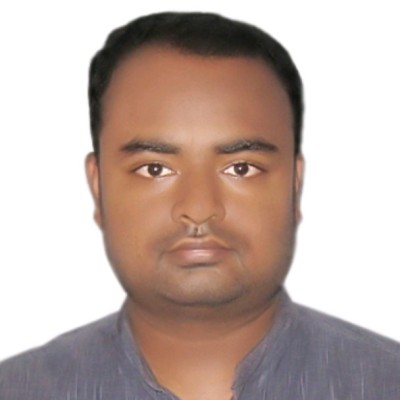}}]{Suman Kumar}
received his Master's degree in VLSI Design \& Embedded Systems from
Delhi Technological University, India, in 2020.
He is currently working toward the Ph.D. degree
under the supervision of Dr. Anup Das from the
Department of Electrical and Computer Engineering,
Drexel University, Philadelphia. His research interests
include neuromorphic computing and the spiking
neural network.
\end{IEEEbiography}

\begin{IEEEbiography}[{\includegraphics[width=1in,height=1.25in,clip,keepaspectratio]{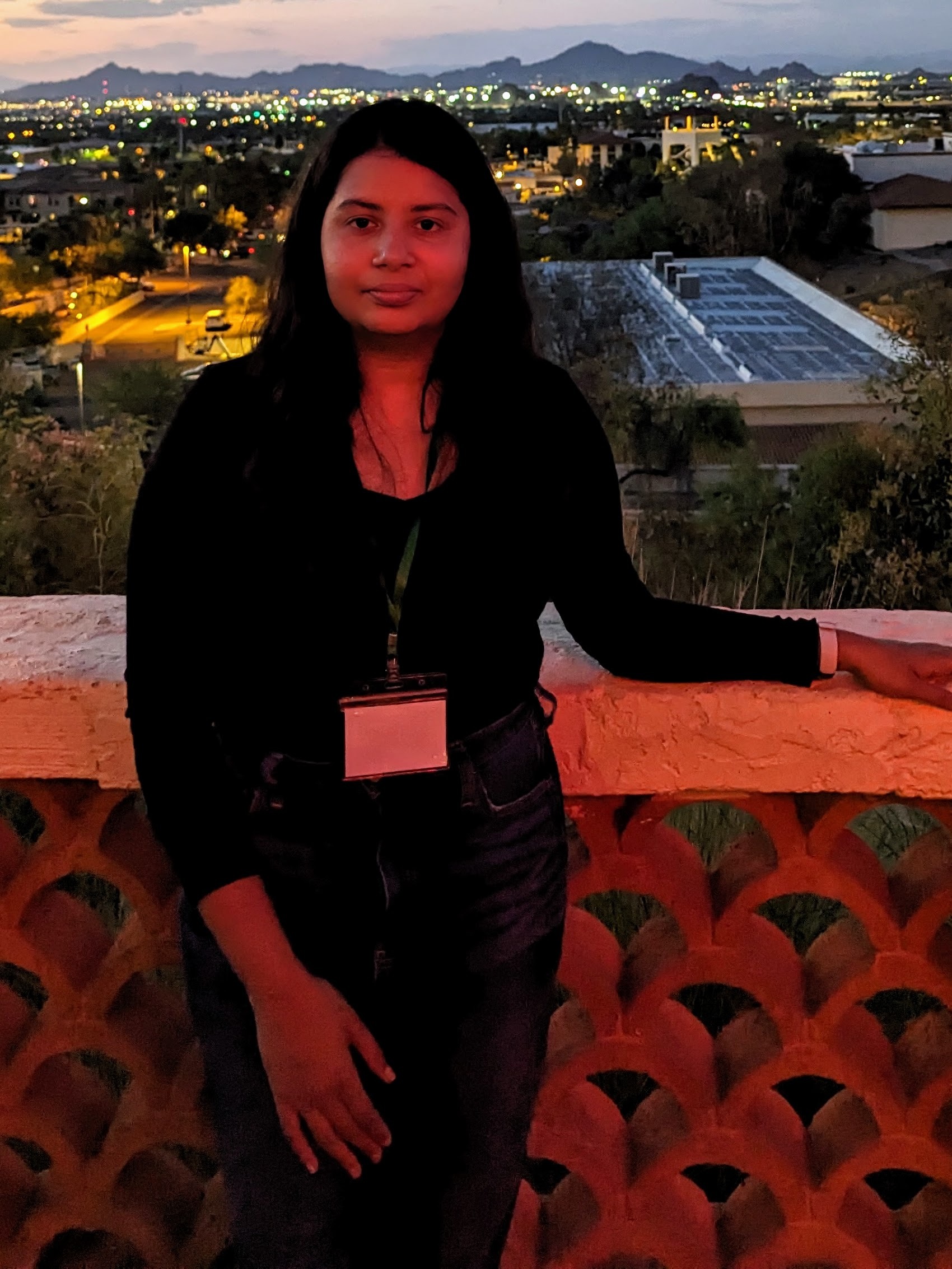}}]{M. Lakshmi Varshika}
earned a B.E. degree from Birla Institute of Technology and Science, Pilani, India, in 2018, and an MSc from the University of California, Santa Barbara, in electrical and computer engineering in 2021. She is currently working towards her PhD degree at Drexel University in Philadelphia. Her research interests include the development of efficient hardware architectures for spiking neural network-based neuromorphic accelerators. Additionally, she is involved in designing neuromorphic circuit solutions for accelerators. Lakshmi Varshika Mirtinti has received the Best Research Video Award at DAC in 2021 and 2022, and she is a finalist for the Best Paper Award at MWSCAS 2023.
\end{IEEEbiography}

\begin{IEEEbiography}[{\includegraphics[width=1in,height=1.25in,clip,keepaspectratio]{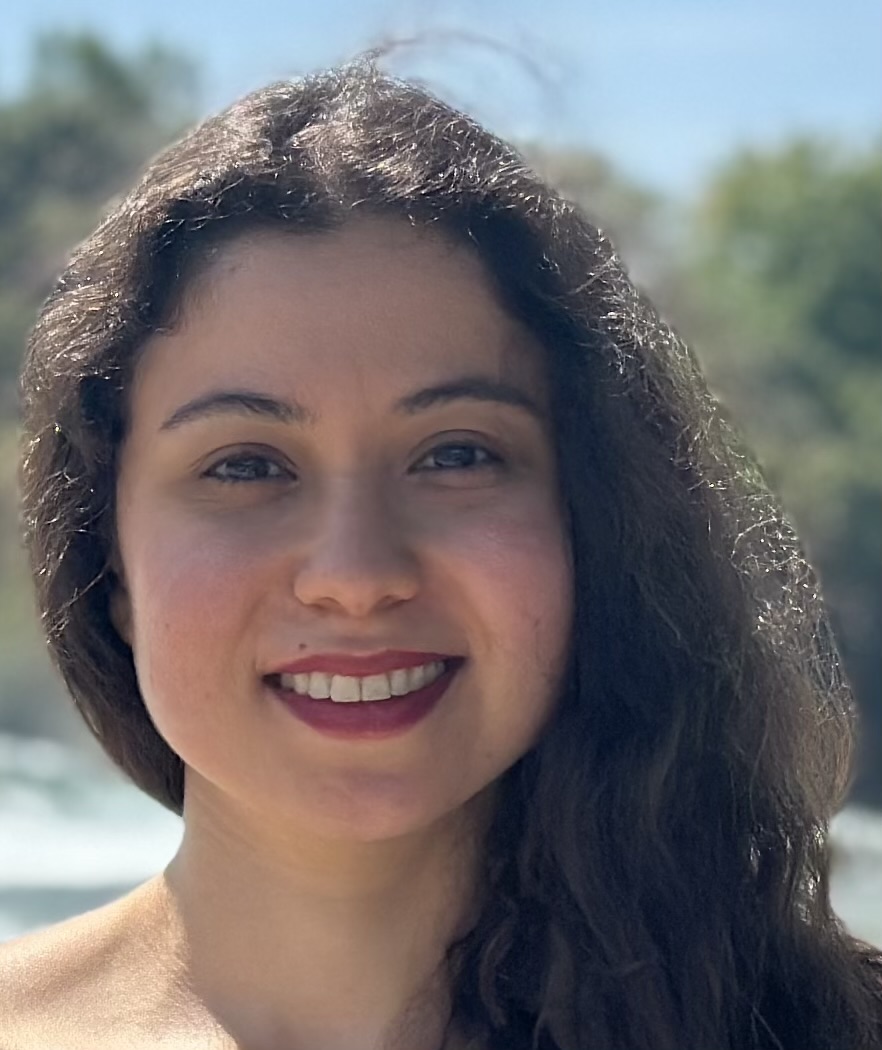}}]{Arghavan Mohammadhassani}
is currently a Ph.D. student at Drexel University. Her focus of research is developing and optimizing memory/storage systems for modern heterogeneous multicore systems, which consist of a variety of cores such as CPUs, GPUs, and Neuromorphic Processing (NP) cores.
\end{IEEEbiography}

\begin{IEEEbiography}[{\includegraphics[width=1in,height=1.25in,clip,keepaspectratio]{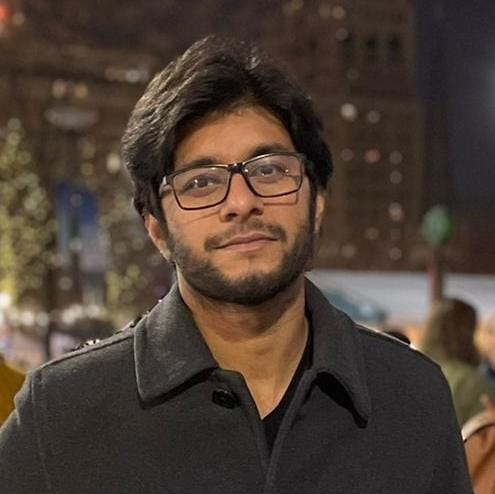}}]{Abhishek Kumar Mishra}
is a Ph.D. candidate at Drexel University. His research focuses on machine learning, neuromorphic computing, and natural language processing. He received his MS degree from SUNY Buffalo in 2020.
\end{IEEEbiography}

\begin{IEEEbiography}[{\includegraphics[width=1in,height=1.25in,clip,keepaspectratio]{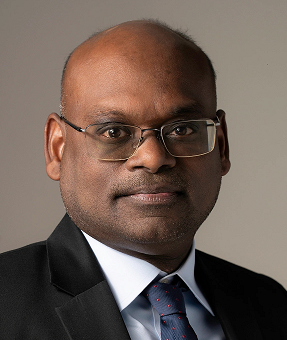}}]{Nagarajan Kandasamy}
is a Professor in the Department of Electrical and Computer Engineering at Drexel University. His current research interests are in the areas of computer architecture, parallel processing, and embedded and real-time systems. He received his Ph.D. in Computer Science and Engineering from the University of Michigan, Ann Arbor. He is a senior member of the IEEE.  
\end{IEEEbiography}

\begin{IEEEbiography}[{\includegraphics[width=1in,height=1.25in,clip,keepaspectratio]{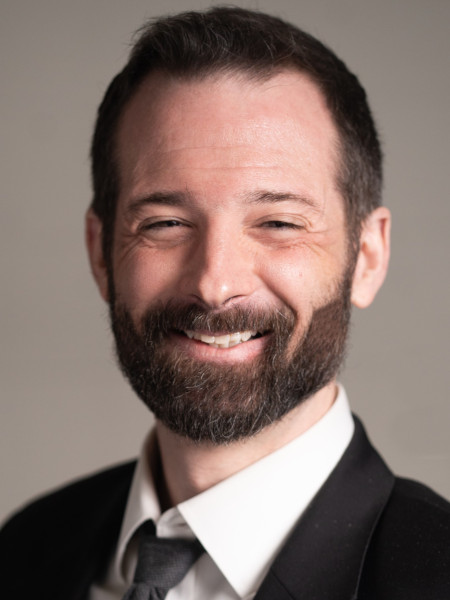}}]{James~Shackleford}
(S’03–M’13) received the B.S. degree in electrical engineering from Drexel University, Philadelphia, PA, USA, in 2006, and the M.S. degree in electrical engineering and the Ph.D. degree in computer engineering from Drexel University, in 2009 and 2011, respectively. He was a Postdoctoral Fellow of the Radiation Oncology Department, Massachusetts General Hospital, Boston, MA, USA, in 2012 and an Adjunct Assistant Professor of medicine with the Perelman School of Medicine, University of Pennsylvania, Philadelphia, PA, USA, from 2017 to 2020. He is currently an Associate Professor of computer engineering with Drexel University, where he now serves as the Associate Department Head for Undergraduate Affairs.  His research has been concerned with computer vision algorithm development for medical image processing with application to image-guided radiotherapy.
Dr. Shackleford is a member of the ACM. 
\end{IEEEbiography}


\begin{IEEEbiography}[{\includegraphics[width=1in,height=1.25in,clip,keepaspectratio]{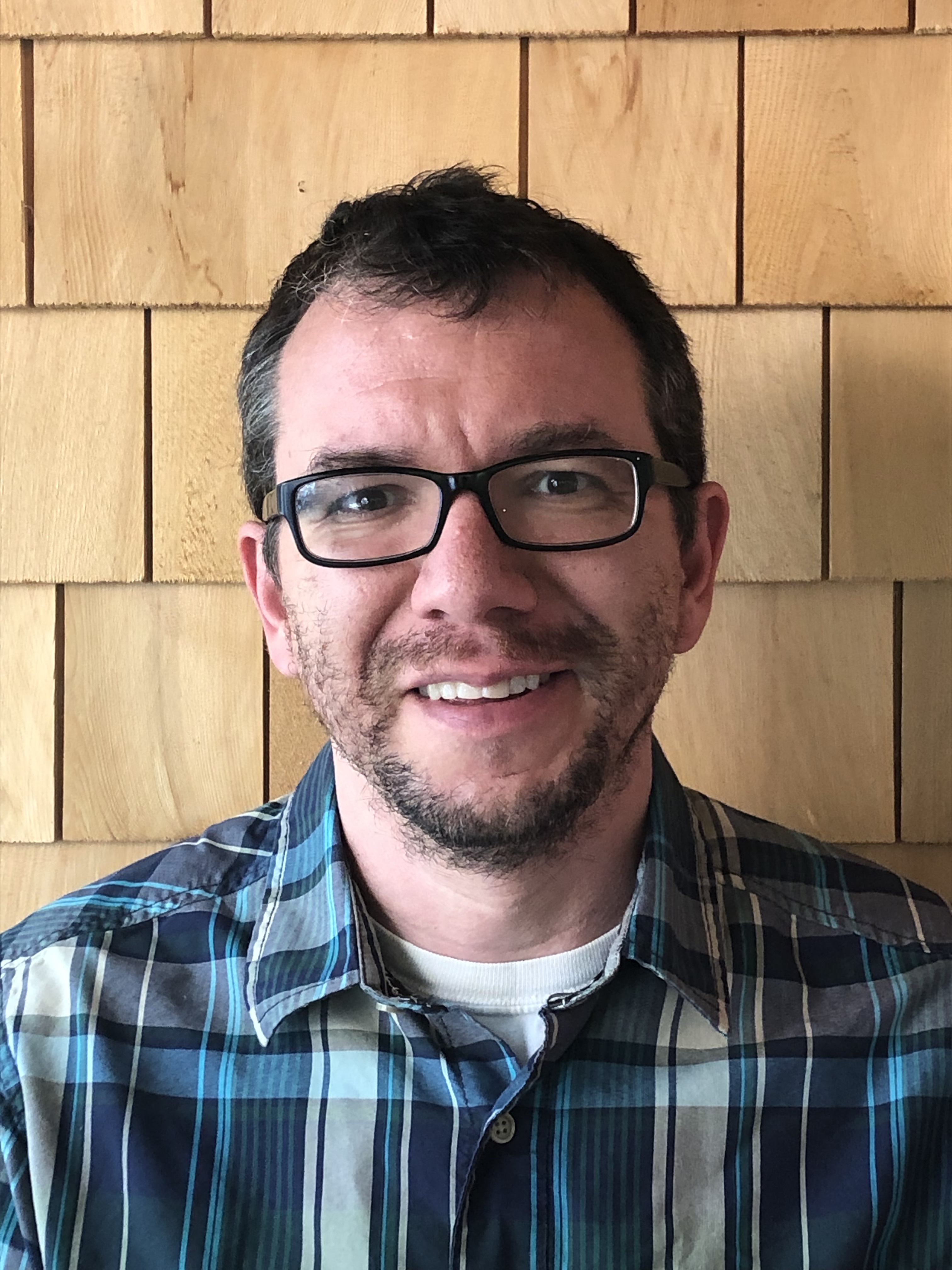}}]{Eric Gallo}
is a Technical R\&D Senior Principal at Accenture Labs. His current research interests are in areas of low power edge processing, sustainable electronics and smart materials. He received his PhD in Electrical and Computer Engineering from Drexel University, Philadelphia PA in 2011. 
\end{IEEEbiography}

\begin{IEEEbiography}[{\includegraphics[width=1in,height=1.25in,clip,keepaspectratio]{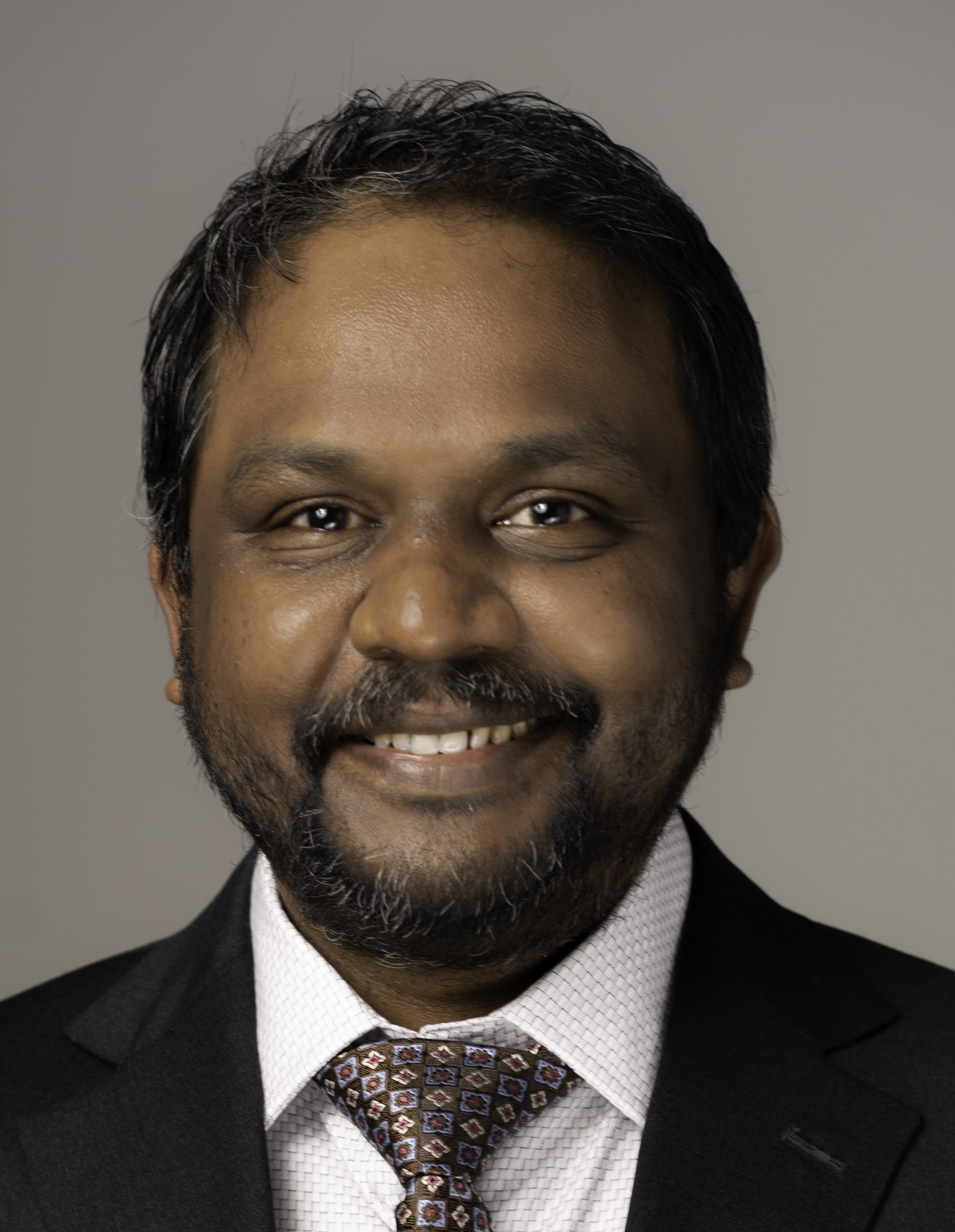}}]{Anup Das}
is an Associate Professor at Drexel University and Associate Department Head for Graduate Studies. He received a Ph.D. in Embedded Systems from the National University of Singapore in 2014.  Following his Ph.D., he was a postdoctoral fellow at the University of Southampton and a researcher at IMEC. His research focuses on neuromorphic computing and architectural exploration. 
He received the United States National Science Foundation CAREER Award in 2020 and the Department of Energy CAREER Award in 2021 to investigate the reliability and security of neuromorphic hardware.
He is a senior member of the IEEE and a member of the ACM.
\end{IEEEbiography}

\vfill

\end{document}